\documentclass[review]{elsarticle}

\pdfoutput=1

\usepackage{fullpage}
\usepackage{amssymb}
\usepackage{amsmath}
\usepackage{color}
\usepackage{graphicx}
\usepackage{caption}
\usepackage{subcaption}
\usepackage{amsmath}
\usepackage{algpseudocode}
\usepackage{algorithm}
\usepackage[titletoc,title]{appendix}

\usepackage{booktabs}
\usepackage[para,online,flushleft]{threeparttable}
\usepackage{color}
\usepackage[table,xcdraw]{xcolor}
\definecolor{light-gray}{gray}{0.85}

\journal{ }

\biboptions{authoryear}

\begin{document}

\begin{frontmatter}

\title{Learning to Allocate Limited Time to Decisions with Different Expected Outcomes}

\author{Arash Khodadadi\fnref{mycorrespondingauthor}}
\author{Pegah Fakhari}
\author{Jerome R. Busemeyer}
\address{Indiana University, Department of Psychological and Brain Sciences, Bloomington, IN, United States}
\fntext[mycorrespondingauthor]{Corresponding author. Indiana University, Department of Psychological and Brain Sciences, 1101 E. 10th street, 47405-7007, Bloomington, IN, United States.}

\begin{abstract}
The goal of this article is to investigate how human participants allocate their limited time to decisions with different properties. We report the results of two behavioral experiments. In each trial of the experiments, the participant must accumulate noisy information to make a decision. The participants received positive and negative rewards for their correct and incorrect decisions, respectively. The stimulus was designed such that decisions based on more accumulated information were more accurate but took longer. Therefore, the total outcome that a participant could achieve during the limited experiments' time depended on her ``decision threshold", the amount of information she needed to make a decision. In the first experiment, two types of trials were intermixed randomly: hard and easy. Crucially, the hard trials were associated with smaller positive and negative rewards than the easy trials. A cue presented at the beginning of each trial would indicate the type of the upcoming trial. The optimal strategy was to adopt a small decision threshold for hard trials. The results showed that several of the participants did not learn this simple strategy. We then investigated how the participants adjusted their decision threshold based on the feedback they received in each trial. To this end, we developed and compared 10 computational models for adjusting the decision threshold. The models differ in their assumptions on the shape of the decision thresholds and the way the feedback is used to adjust the decision thresholds. The results of Bayesian model comparison showed that a model with time-varying thresholds whose parameters are updated by a reinforcement learning algorithm is the most likely model.
In the second experiment, the cues were not presented. We showed that the optimal strategy is to use a single time-decreasing decision threshold for all trials. The results of the computational modeling showed that the participants did not use this optimal strategy. Instead, they attempted to detect the difficulty of the trial first and then set their decision threshold accordingly.
\end{abstract}


\end{frontmatter}

\section{Introduction}
Suppose you are taking an exam. You have one hour to answer as many questions as you can. In addition, suppose that there are two types of questions, easy and hard. How much time should you spend on each question? For example, if the questions are presented sequentially and the first question is hard, would you be willing to spend 10 minuets on that question? In this scenario, every moment that one spends on one question, less will remain for other questions and so fewer questions can be answered in limited time. On the other hand, by answering the questions too fast, the accuracy drops and one may be able to answer only a few questions correctly. This results in a trade-off between the speed and the accuracy.

This is an example of a more general problem in which a living organism has to allocate a limited resource to different courses of actions. Some examples of a limited resource are: energy, time, memory, attention and so on. Usually, spending more of the resource on a course of action results in more desirable outcome for those actions. However, spending more of the resource on some actions leaves less for the other actions, and this may result in lower \textit{total outcome}. Therefore, the organism must allocate the resource ``wisely" to obtain the maximum total outcome from all actions. 

A situation in which this sort of trade-off arises naturally is perceptual decision making in which the animal has to make decisions based on noisy information. Usually, by spending more time the animal can make more accurate decisions which in turn lead to more desirable outcomes. A large amount of research has focused on explaining the relationship between the decision time and the accuracy in perceptual decision making, both theoretically and experimentally  \cite{ratcliff_theory_1978,townsend_stochastic_1983,ratcliff_connectionist_1999,smith_stochastic_2000,usher_time_2001,brown_ballistic_2005,gold_banburismus_2002,kiani_bounded_2008,teodorescu_disentangling_2013,jones_unfalsifiability_2014,khodadadi_mimicry_2015}. The most popular theoretical framework for explaining the mechanism underlying this relationship is provided by a class of models known as \textit{sequential sampling models}. A common assumption between different instantiations of these models is that the animal sequentially samples evidence favoring each of the possible decisions. Since these samples are noisy, a decision based on one sample will be very inaccurate. Instead, the brain accumulates these samples until the accumulated evidence favoring one of the decisions reaches a specific level called the \textit{decision threshold}. Larger values of the decision threshold lead to slower but more accurate decisions. The rate at which the information is accumulated is proportional to the difficulty of the stimulus and so it is controlled by the experimenter and not the participant. 

Experimental results together with computational modeling have shown that human participants adjust the value of their decision threshold in response to the emphasis on the speed or the accuracy in the instructions of the experiment \cite{luce_response_1986,ratcliff_diffusion_2002,wagenmakers_diffusion_2008,forstmann_cortico-striatal_2010}. This experimental paradigm, provides evidence that human participants can adjust their decision threshold when they are asked to do so. However, it does not show if this threshold adjustment will occur in order to maximize the outcome. Recently, some theoretical and experimental work has investigated this question. Gold and Shadlen \cite{gold_banburismus_2002} proposed an experimental paradigm in which the participants had to make a sequence of decisions during a limited time. The participants received some rewards or punishments for their correct or incorrect decisions. Since time is limited, the participant should balance between her speed and accuracy to achieve the maximum amount of reward during the experiment. Bogacz and his colleagues \cite{bogacz_physics_2006} investigated the optimal strategies in this paradigm. Specifically, they showed the relationship between the optimal value of the decision threshold and the parameters of the experiment including the difficulty of the stimulus and the value of the reward and punishment. More recently, \cite{simen_reward_2009} and \cite{balci_acquisition_2011} examined experimentally if human participants can learn the optimal decision threshold in this paradigm.

These studies have shed light on several aspects of the decision making mechanisms involved in balancing between speed and accuracy in information accumulation paradigms. However, many questions have remained unanswered. In this paper, we extend the previous research in several directions in order to investigate some of these questions. We outline these directions next.

\subsection{A novel stimulus and decision paradigm}
The speed-accuracy trade-off have been mainly investigated using perceptual decision making paradigms. These experiments are appealing because it is easy to manipulate the difficulty of the task to achieve a wide range of accuracy (from chance level to perfect accuracy) and reaction time. However, using these stimuli for studying the properties of the decision thresholds have several drawbacks. First, 
in the tasks which are commonly used to study perceptual decision making, for example the random dot motion experiment \cite{britten_analysis_1992,shadlen_neural_2001}, neither the accumulated information nor the decision threshold are directly observable. The only observable variables are the participants' choice and reaction time in each trial. Therefore, to infer the properties of the decision threshold in these experiments, one should either use the neuro-physiological data \cite{shadlen_neural_2001,ratcliff_dual_2007,kiani_bounded_2008,ivanoff_fmri_2008,forstmann_cortico-striatal_2010}, or computational modeling \cite{ratcliff_theory_1978,smith_psychophysically_1995,ratcliff_comparison_2004,usher_time_2001}. This makes the inference about the properties of the decision thresholds harder than if the decision threshold could be observed directly. Second, for the same level of task difficulty, there is usually a large amount of variations in the participants' performance. This is due to individual differences in perceiving the same stimulus. In the language of the sequential sampling models, for the same stimulus, the participants have different rate of information accumulation. For this reason, the properties of he optimal decision threshold will be different for different participants. Third, there is usually a large perceptual learning effect in these tasks. With experience, the rate of information accumulation increases for a participant. Therefore, the properties of the optimal decision threshold changes for a participant during the experiment. Fourth, the participants' average reaction time in these experiments is usually very short. As we will argue later, this may put some constraints on the shape of the decision thresholds.

To address these issues, we propose a new stimulus and decision paradigm. Using this stimulus, we are able to observe the decision threshold and the rate of information accumulation in each trial directly. Also, since the rate of information accumulation is controlled by the experimenter, the optimal decision threshold will be the same for all participants and will not change during the experiment.  

\subsection{Allocation of time to decisions with different properties}
In all aforementioned studies on speed-accuracy trade-off, it is assumed that the participants adopt only one decision threshold in all trials and they adjust it during the experiment. In particular, in each block of these experiments, all trials had the same level of difficulty. Also, the reward and punishment associated with the correct and incorrect responses, were the same for all trials in a block. Therefore, to be optimal, the participant needs to set only one decision threshold for all trials. However, as our first example suggests, in many real life situations, one needs to allocate time between decisions with different properties. Not only the difficulty of the decisions may differ (as in the exam example mentioned above), but also their expected outcomes may differ. Sometimes, harder decisions are associated with higher stakes. For example, publishing a paper in a higher-impact journal has larger outcome, but the likelihood of rejection is also higher. The reverse is also the case in many situations. For example, for an animal seeking for food, finding a larger fruit is easier and its outcome is obviously higher. When the animal has to allocate time between decisions with different properties, it may be optimal to set different decision thresholds for each type of decisions. Behaving optimally in these situations is much harder because to know how much time should be spent on one decision, one should have an estimate of the expected outcome of other decisions.

In this paper, we extend the aforementioned experimental design to situations in which the participants have to make a sequence of decisions with different properties. Experiments A reported below consists of 40 blocks of trials. The blocks duration is fixed (one minute) and so the number of trials in each block depends on the participant's speed in responding in each trial. The participants receive positive (negative) reward for their correct (incorrect) decisions in each trial. Each trial could be either an ``easy" or ``hard" trial. Obviously, in the easy trials, detecting the correct response is easier than the hard trials. In addition, the easy trials are associated with higher absolute values of both positive and negative reward. We will show the relationship between the experiments' parameters (including the values of the rewards, the difficulty level of the trials and so on) and the average time that should be spent on each trial in order to achieve the maximum amount of expected total outcome. We chose the values of the experiments' parameters such that to be optimal (i.e., to achieve the maximum possible total outcome) the participant must adopt a much lower value of the decision threshold for the hard trials than the easy trials.

\subsection{Computational models of threshold adjustment}
Previous research on rewarded perceptual decision making experiments have mainly focused on the ``optimal speed-accuracy trade-off", specifically, quantifying the optimal behavior and examining if the participants can learn this optimal behavior \cite{bogacz_physics_2006,frazier_sequential_2008,simen_reward_2009,balci_acquisition_2011,karsilar_speed_2014,khodadadi_learning_2014}. However, little is known about \textit{how} the participants adjust their decision threshold in these experiments. Our main focus in this paper is on this question. We attempt to develop computational models that can explain how the participants adjust their decision threshold after receiving the feedback in each trial.

We will consider several computational models which differ in their assumptions on two dimensions: first, the shape of the decision thresholds, and second, how the feedback is used to adjust the decision thresholds. We explain each of these briefly next.

An important aspect of any computational model for information accumulation experiments is its assumption about the form of the decision threshold. Traditional sequential sampling models assume that the decision threshold remains constant within a trial. More recently, researchers have considered models with decision thresholds that change as a function of the elapsed time in a trial \cite{ditterich_stochastic_2006,frazier_sequential_2008,churchland_decision-making_2008,drugowitsch_cost_2012,zhang_time-varying_2014,khodadadi_mimicry_2015}. Specifically, the time-varying decision thresholds have been shown to be optimal in several experimental designs. To investigate the form of the decision thresholds used by the participants in our experiments, we will consider two implementations of each of the computational models: one with time-constant decision threshold and one with time-varying decision threshold. Interestingly, our results showed that the time-varying version of each model fitted better than the time-constant version. 

To model the mechanisms for adjusting the decision thresholds, we consider four categories of models: baseline models in which no learning occurs, models that adjust threshold only based on the rewards and ignore the time, models which assume the participant tries to achieve desired levels of accuracy and reaction time, and the reinforcement learning models.

Perhaps the reinforcement learning (RL) theory is the most commonly computational framework for describing the learning mechanisms in value-based decision making studies \cite{sutton_reinforcement_1998,daw_reinforcement_2003,dayan_decision_2008}. In this framework, if the total outcome depends only on the rewards in each trial, the experiment can be modeled as a Markov decision process (MDP). This is the most common application of the RL theory in modeling the learning behavior of human participants \cite{odoherty_dissociable_2004,daw_model-based_2011,dezfouli_actions_2013}. In our experiments, however, the total outcome depends on both the rewards and the reaction times in each trial. Therefore, the MDP framework is not appropriate for our application. Instead, we will model our experiments as semi-Markov decision processes (SMDP), an extension of the MDPs in which it is possible to incorporate the effect of the decision times in the total outcome. We then develop RL algorithms for adjusting the decision thresholds in the corresponding SMDPs. The results of comparing the models showed that, although there are individual differences, the RL models provide the best fit to the data of most of the participants in Experiment A.

\subsection{Time at which decision thresholds are set}
Another important aspect that affects the performance in the speed-accuracy trade-off paradigms is the time at which the decision thresholds are set. Conventional sequential sampling models assume that the participants set their decision threshold at the beginning of the trial and before the stimulus is presented. Therefore, in fitting these models to data from experiments in which trials with different levels of difficulty are intermixed and there is no cue for the participants to know the difficulty level of a trial before the stimulus is presented, it is assumed that the participants use the same decision threshold for all trials \cite{ratcliff_connectionist_1999,ratcliff_diffusion_2002,ratcliff_comparison_2004}.

In our Experiment A, a cue presented at the beginning of each trial would indicate the condition (easy or hard) of the upcoming trial. Therefore, the participant could set two different decision thresholds for the trials from the two conditions. To test the assumption that the participants set their decision threshold at the beginning of the trial, in Experiment B the cues were not presented. Other parameters of this experiment were similar to Experiment A. If this assumption is valid, the participants may use two different thresholds for the two types of trials in experiment A, but they must use the same threshold for all trials in Experiment B. We will show that the optimal strategy is to use two different time-constant thresholds for easy and hard trials in Experiment A, and one time-decreasing threshold for all trials in Experiment B. Another possibility, however, is that the participants in Experiment B use a mechanism to detect the difficulty of the trial (by observing the stimulus for some time after the beginning of the trial) and then set the decision threshold accordingly. We will develop computational models to capture each of these strategies and compare them to investigate which strategy can describe the participants' data better. The results of model comparison provided strong evidence for the assumption that the participants try to detect the difficulty of the trials first and set the decision threshold accordingly. 

\section{Material and Methods}
\subsection{Behavioral experiments}

\begin{figure}[!h]
	\centering	
	\includegraphics[width=.5\linewidth]{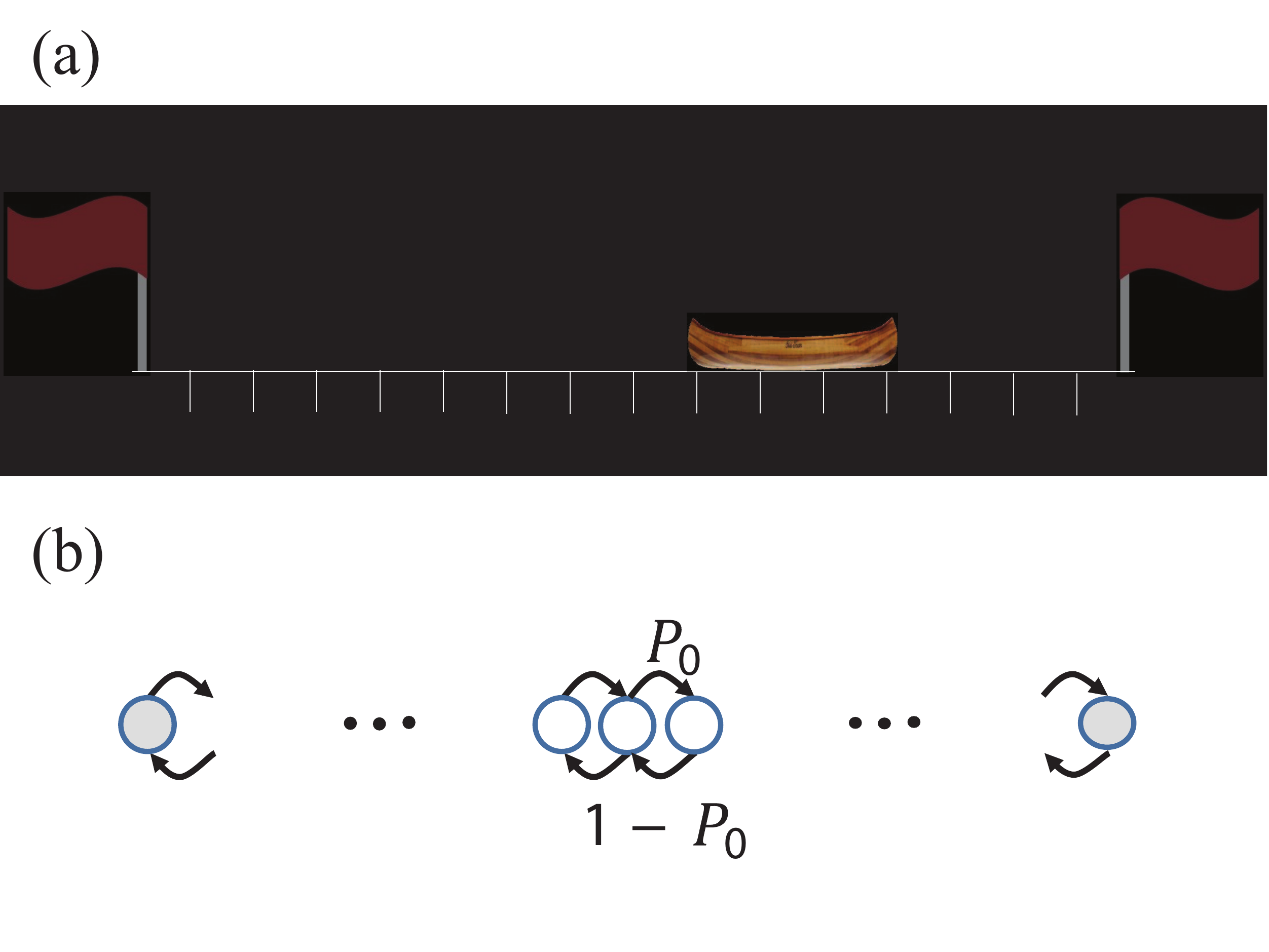}		
	
	\caption{{\bf Canoe movement detection task} (a) Example screen (b) Markov chain governing the canoe movement in each trial. In this figure, the correct direction is right. At each time step, the canoe moves to right with probability $P_0$ and to left with probability $1-P_0$.} \label{fig:Canoe_screen}
\end{figure}

For both experiments reported below, we used a novel stimulus. This stimulus was inspired by this video: . At the beginning of each trial, a canoe is shown at the center of the screen (panel (a) in Figure \ref{fig:Canoe_screen}). After the trial begins, the canoe moves back and forth, to the right and left. The participants were told that the canoe will eventually reach one of the two flags on the right and left. In each trial, the participant must watch the canoe for a while and decide which direction the canoe will eventually go. We call this task the \textit{canoe movement detection task}.

The movement of the canoe was governed by a Markov chain with probability $P_0$. To explain this, suppose that in a trial the ``correct" direction is right. After the trial starts, for $\Delta t$ msec, the canoe will move to the right with probability $P_0$ or moves to the left with probability $1-P_0$. Again, during the next $\Delta t$ msec the canoe moves to the right or left with the same probabilities. The canoe continues moving in this manner until the participant makes her decision. The Markov chain corresponding to this movement is shown in panel (b) of Figure \ref{fig:Canoe_screen}. In experiments below we set $\Delta t=500$ msec which resulted in smooth movement of the canoe. In each trial, the participants responded by pressing the ``c'' or ``m'' keys on the keyboard if they decided that the canoe is moving to the left or right, respectively. The stimulus was presented on 17" Dell monitors controlled by Dell Optiplex 990 computers using the Psychophysics Toolbox in MATLAB \cite{brainard_psychophysics_1997}.

In contrast to the experiments which are commonly used to study perceptual decision making, in the current experiment, both the accumulated information and the decision threshold are observable. At any moment within a trial, the only relevant information to make a correct decision (detecting the correct direction of the canoe) is the position of the canoe at that moment. This is because the canoe movement is governed by a Markov chain and the probability of moving to right or left at any moment is independent of the canoe position and the elapsed time. Therefore, the canoe position at any moment is equivalent to the accumulated information in a sequential probability ratio test. In the sequential sampling models, it is assumed that the participant responds whenever the accumulated information exceeds a decision threshold. Similarly, we assume that in each trial of our experiment the participant responds whenever the canoe position deviates from the center of the screen (the initial position of the canoe) by a specific amount. Based on this assumption, in each trial, the canoe position at the time that the participant responds is the decision threshold in that trial.    

\begin{figure}[!h]
	\centering	
	\includegraphics[width=.3\linewidth]{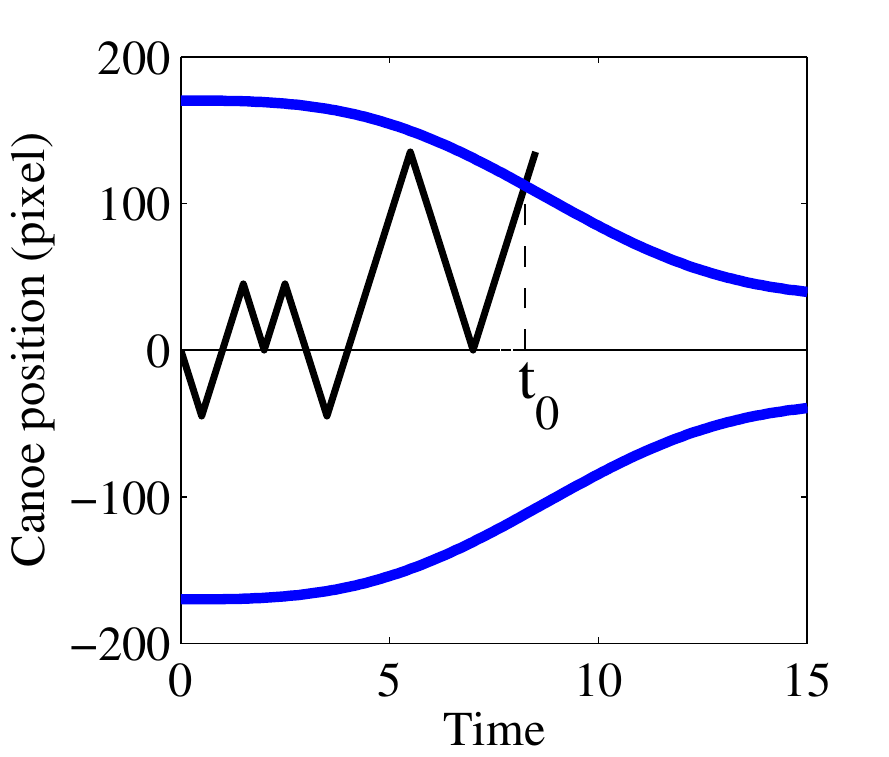}		
	
	\caption{{\bf Sample path of canoe movement.} The black curve shows the sample path of the canoe movement in a trial and the blue curves are examples of time-varying decision thresholds. The upper and lower thresholds correspond to the ``right" and ``left" responses, respectively. In this example, the canoe position has reached the participant's ``right" decision threshold at $t_0=8.2$ secs.} \label{fig:Canoe_SamplePath_DecTh}
\end{figure}

In this paper, we will consider both time-constant and time-varying decision thresholds. Figure \ref{fig:Canoe_SamplePath_DecTh} shows a sample path of the canoe position in a trial and an example of time-decreasing thresholds. In this figure, the horizontal axis is the elapsed time in a trial and the vertical axis is the canoe position, in pixels, relative to the center of the screen. Positions right to the center are considered as positive. As it can be seen, the canoe position has first reached the upper decision threshold at $t_0=8.2$ secs when the canoe position was at 112 pixels. Of course, we cannot observe the whole form of the decision threshold in the trial. Instead, the position of the canoe at the time the participant made her decision in a trial, gives us the value of the decision threshold at that time. We recorded the participants' choice (left or right), the decision time and the value of the decision threshold at the decision time for all trials. In later sections, we will show how we can use these data to infer the shape of the decision thresholds that a participant has used. 

\begin{figure}[!h]
	\centering	
	\includegraphics[width=.2\linewidth]{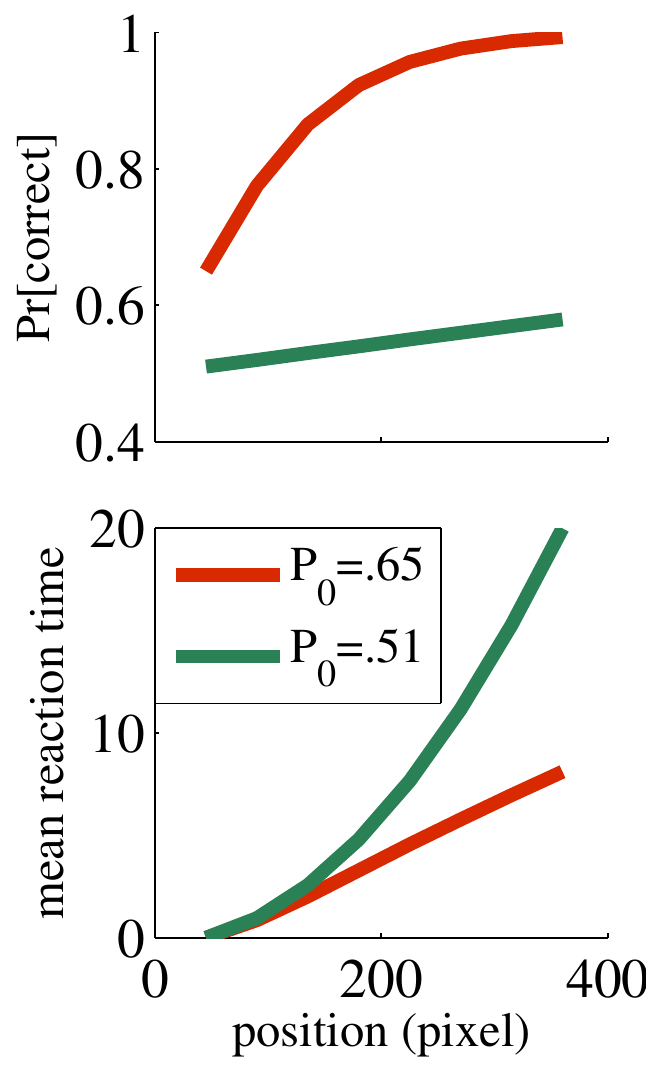}		
	
	\caption{{\bf Accuracy and mean RT in canoe movement detection task.} Top: Accuracy, Bottom: mean RT as functions of the decision threshold for two values of $P_0$.} \label{fig:Canoe_MeanRT_Acc_vs_Thd}
\end{figure}

Since the canoe position is equivalent to the accumulated information, the difficulty of a trial is determined by the value of $P_0$: for large values of this parameter, the canoe moves more consistently toward the correct direction and so the canoe goes to the correct end of the screen faster. Figure \ref{fig:Canoe_MeanRT_Acc_vs_Thd} shows the probability of responding correctly (accuracy) and the mean reaction time (RT) as a function of the decision threshold for two values of $P_0$. To generate these figures, we used time-constant decision thresholds with the same absolute value for the left and right responses.  The horizontal axis in these figures is the absolute value of the decision threshold. For any given value of $P_0$ and the decision threshold, we used the method in \cite{diederich_simple_2003} to compute the accuracy and the mean RT\footnote{Since the canoe movement is governed by a Markov chain, the problem of finding the mean RT and accuracy for a given value of the decision threshold is equivalent to the problem of finding the probability and time of reaching an absorbing state in a Markov chain. There are numerical methods for solving this problem. We used the matrix method in \cite{diederich_simple_2003}.}. Two points should be noted in these figures: First, for all values of $P_0$ when the value of the decision threshold increases, both accuracy and mean RT increase. Therefore, the participants can balance between their speed and accuracy by adjusting their decision threshold. Second, for the same value of the decision threshold, the accuracy is higher for larger values of $P_0$. Therefore, the level of the difficulty is determined by the value of $P_0$.

Here, we should emphasize on another advantage of using this stimulus for the purposes of the current paper: For a given value of $P_0$, the accuracy and the mean RT are determined only by the decision thresholds. In other words, if two participants use exactly the same decision thresholds, their accuracy and mean RT will be the same. This is not the case in the conventional stimuli used in the perceptual decision making tasks. For example, consider the random dot motion task. The difficulty of the task is determined by the ``motion coherence", the percentage of dots that move coherently toward the correct direction. Since different participants have different perceptual ability in detecting the motion, for a fixed level of difficulty, even if two participants use the exact same decision thresholds, their accuracy and/or mean RT may not be the same. More importantly, it is possible that due to perceptual learning the ability of an individual participant in detecting the correct direction of the motion increases by experience during the experiment \cite{law_reinforcement_2009}. Therefore, the participants' performance is a function of both perceptual and decisional processes. In this paper, we are interested in investigating the decisional processes and so it is appealing to use a stimulus which makes the performance only a function of the decisional processes.

We used this stimulus in two experiments. The details of these experiments are given in the next two sections. All studies reported below, were approved by the Indiana university IRB and all participants provided informed written consent before participating in the study.

\subsubsection{Experiment A}
The purpose of this experiment was to investigate how the participants adjust their decision threshold, when they have to allocate limited time between decisions with different properties. We used the canoe movement detection task introduced above. The experiment consisted of 40 blocks of trials. All blocks for all participants were 1 minute long. The number of trials in each block depended on the participant's speed in responding. Each trial was drawn with probability 0.5 from one of the two possible conditions: easy or hard. In the easy trials, the probability that the canoe would move toward the correct direction was $P_0=0.65$. This probability was  $P_0=0.51$ for the hard trials. In the easy trials, the participants would gain or lose 20 coins for a correct or incorrect response, respectively. This pay-off was $\pm1$ coin in the hard trials. In addition, after an incorrect response in an easy trial, the participant had to wait 3 more seconds before the next trial begins. This \textit{delay penalty} did not exist for the hard trials.

\begin{figure}[!h]
	\centering
	\includegraphics[width=.5\linewidth]{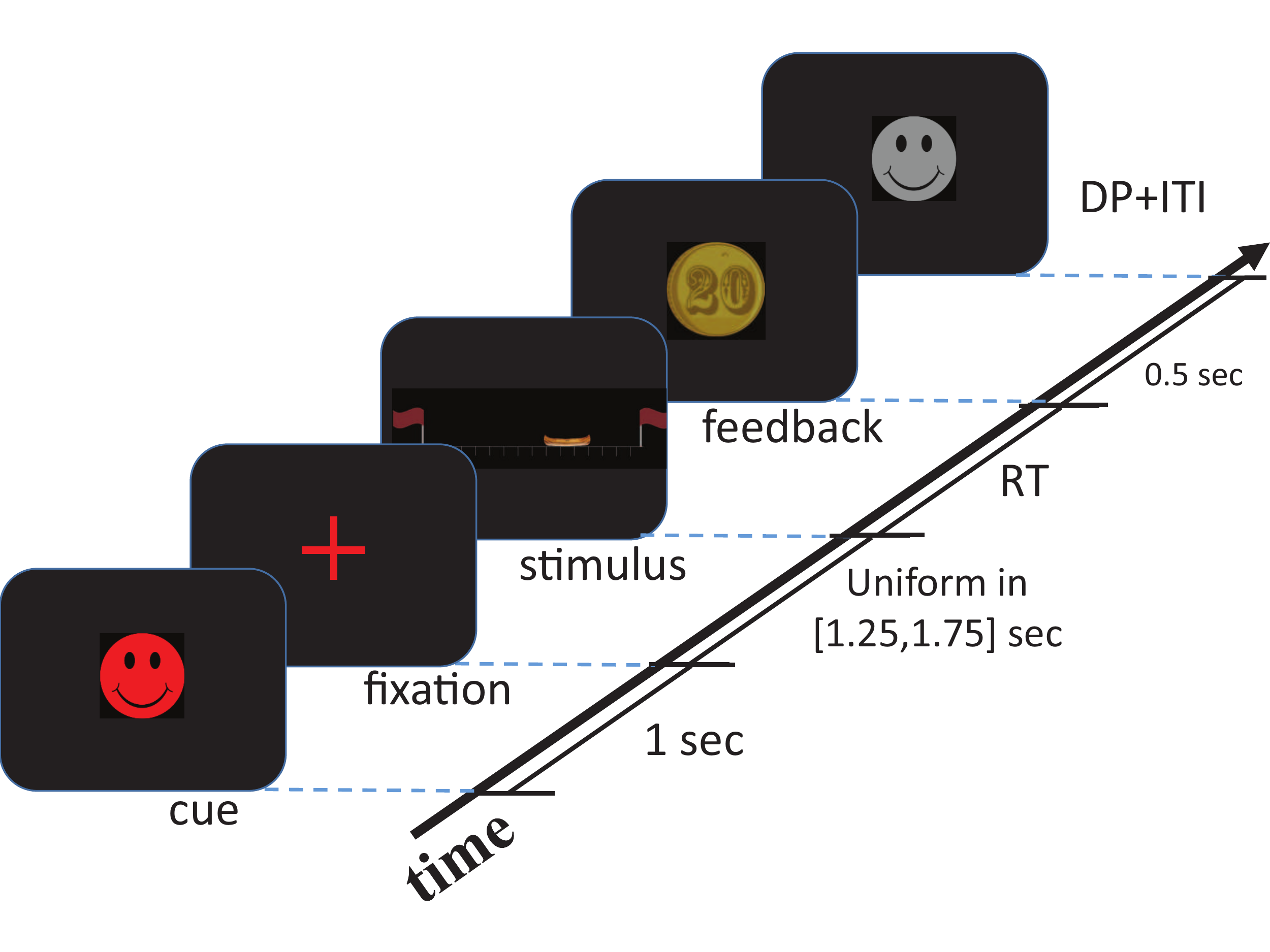}	
	\caption{{\bf Experiment A.} The timeline of events in one trial of Experiment A. See the text for more details. DP: delay penalty, RT: response time, ITI: inter-trial interval.} \label{fig:Canoe_timeline}
\end{figure}

The timeline of one trial is illustrated in Figure \ref{fig:Canoe_timeline}. At the beginning of each trial, a cue was presented that would indicate the condition of the upcoming trial: a red smiley face would indicate an easy trial and a green smiley face would indicate a hard trial. Then a fixation cross-hair was presented for a random time drawn uniformly from interval $[1.25,1.75]$ secs. The stimulus was presented then and remained on the screen until the participant responded. After the participant responded, the feedback was shown for 0.5 secs. After that, a gray smiley face was shown for 1 sec before the next trial started. In the easy trial, if the response was incorrect this time was extended to 4 secs.   

The participants were informed about this structure of the task. Specifically, they were told that since the blocks duration is fixed, to experience more trials they should respond faster, but on the other hand being faster reduces accuracy. Also they were informed that there are two types of trials with different pay-off structure, and that the cues indicate the condition of the upcoming trial. participants were motivated to collect as many coins as they could in the 40 blocks by being told that they will receive $\$1$ for each 1000 coins they collected in the study. A total of 29 participants (age:19-27, 14 female) participated in this study. Three participants were excluded from all analyses because their performance was at the chance level in both easy and hard conditions.

\subsubsection{Experiment B}
This experiment is similar to Experiment A with one crucial difference: no cue is presented at the beginning of the trials. Therefore, the participants could not know what condition a trial is coming from. The pay-off structure for the two conditions was exactly the same as Experiment A. This experiment consisted of 35 one-minute blocks. Also, in the easy trials $P_0=0.75$.

A total of 22 participants (age:19-30, 12 female) participated in this study. Two participants were excluded from the analysis because their performance was at the chance level. 

\subsection{Modeling experiments as semi-Markov decision processes}
The main focus of this paper is on developing and comparing computational models of the participants' threshold adjustment in the above experiments. The RL theory and the accompanying algorithms have been used extensively to model the learning behavior in rewarded decision making experiments. Similar to these experiments, in Experiments A and B the participants should learn to make their decisions in each trial such that the total outcome achieved during the experiment is maximized. In addition, the relationship between the decisions and the achieved outcomes is probabilistic. The RL models have been shown to be powerful tools for explaining the behavior in these situations. It seems reasonable, therefore, to conjecture that the thresholds adjustment in our experiments can be explained well with the RL algorithms.

In experiments in which the total outcome depends only on the rewards achieved in each trial, the first step in constructing an RL model is to describe the experiment as a Markov decision process (MDP). In our experiments, however, the total outcome depends not only on the rewards but also on the decision times in each trial. We, therefore, model the experiments in a framework called semi-Markov decision processes (SMDP) which is a generalization of the MDPs.  To make the paper self-contained we give a brief explanation of the SMDPs next. In the next two sub-sections, we show how the components of Experiments A and B can be mapped to the components of an SMDP. In a later section, we show how the problem of learning the decision thresholds can be modeled in this framework.

An SMDP models the interaction between an \textit{agent} and the \textit{environment}. Since we are using this framework to model the learning behavior of the participants in the experiments, the agent is the participant and the environment is the experiment. An SMDP is specified by a 5-tuple $<S,A,T,R,D>$. The state space $S$, is the set of all possible states in the experiment. Intuitively, a state specifies a unique ``situation" in the experiment, and so the state space includes all possible situations in the experiment. The action space $A$, is the set of all possible actions in all the states. In the SMDP framework, at each step, the environment is in one of the possible states. The agent takes one of the possible actions in that state. After taking the action, the environment remains in the same state for some time and then transitions to a new state, and the agent receives some rewards (positive or negative). The transition between states, the rewards and the time between transitions are all stochastic. Assume that at step $k$, the agent is in state $s_{i,k}$ and takes action $a_k$. The probability of transition from $s_{i,k}$ after taking action $a_k$ to a new state $s_{j,k+1}$ is denoted $T^a_{i,j}=\Pr(s_{j,k+1}|s_{i,k},a_k)$. Similarly, the probability of receiving reward $r_k$ is $R^a_i(r_k)=\Pr(r_k|s_{i,k},a_k)$, and the probability of the transition time being $d_k$ is $D^a_i(d_k)=\Pr(d_k|s_{i,k},a_k).$\footnote{Generally, both the reward and the transition time could depend on the new state. However, as we will see shortly, we do not need this assumption for modeling the experiments in this paper and so we assume that they depend only on the old state and the taken action.}

The probabilities $T^a_{i,j}$, $R^a_i(r)$ and $D^a_i(d)$, specify the \textit{dynamic} of the environment. It is usually assumed that the dynamic of the environment is unknown to the agent. The goal of the agent is to learn the \textit{optimal policy} by interacting with the environment. A policy $\pi_{s,a}$, is a mapping from a state $s$ to the probability of taking each possible action $a$ in that state, that is $\pi_{s,a}=\Pr(a|s)$. Therefore, the goal of the agent is to learn which action to take in each state to achieve the optimal performance. Several measures of optimality have been considered in the literature for SMDP problems. We will adopt the \textit{average reward rate} defined as follows:

\begin{equation}
\rho_\pi(s)=\lim_{N \rightarrow \infty} \frac{E[\sum_{\tau=0}^{N} r_{\tau}]}{E[\sum_{\tau=0}^{N} d_{\tau}]}
\label{eq:Modeling_1}
\end{equation} 

In this equation, $\rho_\pi(s)$ is the average reward rate in state $s$ given that the participant is following policy $\pi$ in all states. In the next section, we will explain why this is an appropriate measure of optimality of performance in our experiments. 

Das et. al., \cite{das_solving_1999} have investigated the optimality in SMDPs with this measure. It has been shown that under some technical assumptions (which hold for our experiments), the average reward rate does not depend on the initial state and will be the same for all states (see \cite{das_solving_1999}, the paragraph after equation 4 for more details). Therefore, we drop the dependency to the state and use $\rho_\pi$ as the average reward rate in all states of the experiment when following policy $\pi$. The optimal value of the average reward rate is $\rho^*=\max_{\pi}\{\rho_\pi\}$. An agent is behaving optimally if its following a policy $\pi^*$ which is yielding the maximum average reward rate, that is $\rho_{\pi^*}=\rho^*$. Theorem 1 in \cite{das_solving_1999} shows that this optimal value is unique and it satisfies the following system of equations:

\begin{equation}
V^*(s_i)=\max_a \bigg \{ \sum_{r}r\cdot R^a_i(r) - \rho^* \cdot \int \tau\cdot D^a_i(\tau) d\tau + \sum_{s_j \in S} T^a_{i,j}\cdot V^*(s_j)  \bigg \} , \forall s_i \in S 
\label{eq:Modeling_2}
\end{equation} 

\noindent where $V^*(s_i)$ is the \textit{optimal value} of state $s_i$. The unknown variables in this system of equations are optimal state values $V^*(s_i)$ and the optimal average reward rate $\rho^*$. The importance of these equations, called the \textit{Bellman optimality equations}, is that by solving them we can determine the optimal policy $\pi^*$: given the values of $V^*(s_i)$, the optimal action in state $s_i$ is $a^*=\arg\max_{a}\{ Q(s_i,a) \}$, where $Q(s_i,a)$ is the term in the braces in Equation \ref{eq:Modeling_2}. When the dynamic of the environment ($T^a_{i,j}$, $R^a_i(r)$ and $D^a_i(d)$) is known to the agent, \textit{dynamic programming} methods can be used to solve these equations and find the optimal policy \cite{bertsekas_neuro-dynamic_1996,sutton_reinforcement_1998}. However, when the dynamic is unknown (which is the case for our experiments as we will see soon), the agent should learn the optimal policy only by interacting with its environment. The class of algorithms for learning the optimal policy in stochastic environment with unknown dynamics are known as RL algorithms.

To model Experiments A and B as SMDPs, we should specify the state and action spaces, and the corresponding distributions $T^a_{i,j}, R^a_i(r)$ and $D^a_i(d)$. In the subsequent sections, we specify these for each of the experiments.

\subsubsection{SMDP model of Experiment A}
As it was mentioned above, the states are distinct possible situations in the environment. In a sense, in Experiment A there are two distinct situations: easy trials and hard trials. Therefore, we can model this experiment with an SMDP with two states. In each trial, the environment (experiment) is in one of these two possible states. Since in this experiment, the two types of trials are presented randomly with equal probability, the transition probabilities are $T^a_{i,j}=0.5$ for $i,j=E,H$ (for easy and hard) and for all actions $a$. In words, the transition probabilities does not depend on the actions.

Given this state space, the specification of the reward and transition time in each state is simple. The reward is the number of coins that the participant receives in each trial. The transition time is the time between the beginning of a trial and the next trial. This time includes the cue, fixation and reward presentation time, the participant's reaction time, and the delay penalty in the incorrect trials.  

Finally, we must specify the set of possible actions. Since in each trial (or state) there are two possible responses, \textit{left} and \textit{right}, one might consider these as the set of possible actions in each state. However, this is not an appropriate choice for actions for the state space that we are considering. The reason is that the actions must be defined such that the probabilities $R^a_i(r)$ and $D^a_i(d)$ are well-defined. Suppose that in a hard trial the participant has chosen the left response. Given this information, one cannot specify either the probability of the reward that the participant will receive, or the time it takes to transition to the next trial. Therefore, we propose another choice for the actions: the value of the decision thresholds. This is an appropriate choice for the action space for two reasons. First, based on the assumption of the sequential sampling models, the value of the decision threshold is set by the participant, and so it is reasonable to model the value of the decision threshold in each trial, as the action that the agent (the participant) has taken\footnote{It is important to note that, since the threshold can take any positive value, the action space defined in this way is continuous.}. Second, given the state (hard or easy) and the value of the decision threshold, we can determine the probability that the response will be correct, as well as the distribution of the reaction time which determines the transition time. In other words, the probabilities $R^a_i(r)$ and $D^a_i(d)$ are well-defined given the state and the action.

\begin{figure}[!h]
	\centering
	\includegraphics[width=1\linewidth]{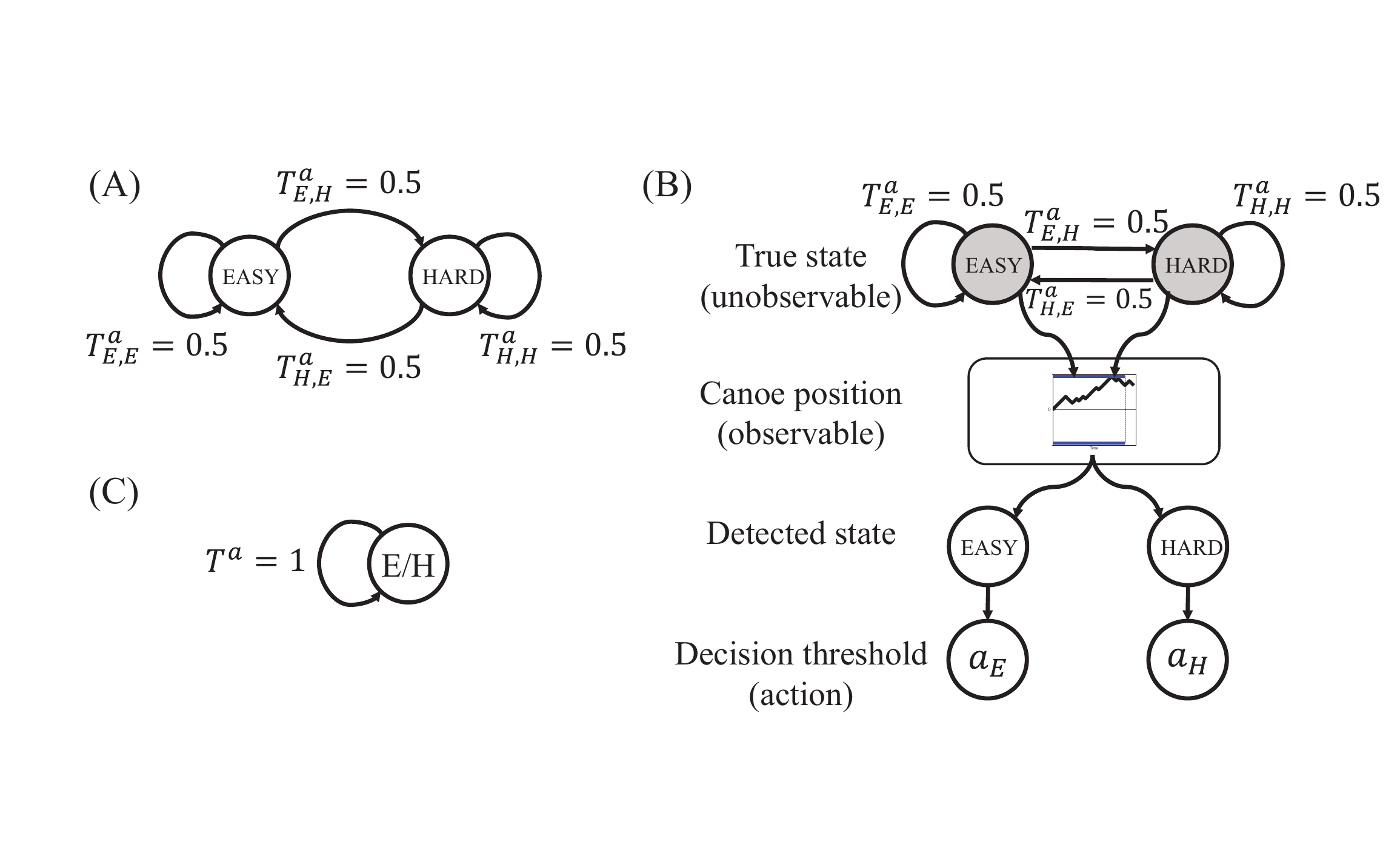}	
	\caption{{\bf SMDPs corresponding to the experiments.} (a) two-state SMDP model of Experiment A. The states correspond to the easy and hard trials. Since these trials are intermixed randomly, the transition probabilities are all equal to 0.5 (shown on the arrows). (b) One-state SMDP model of Experiment B. In this model, all trials are considered as one state and the participant sets one decision threshold for all trials. (c) Two-state SMDP model of Experiment B. In this model, in each trial the true state of the environment is either easy or hard. However, the participant cannot observe this state. Instead, she tries to infer the true state by observing the canoe movement at the beginning of the trial and then sets the corresponding decision threshold. } \label{fig:SMDP}
\end{figure}

In sum, after observing the cue at the beginning of each trial, the participant is in one the two states, easy or hard. She takes the action $a$ which means that she sets the decision thresholds for the \textit{right} and \textit{left} responses at the values $\pm a$\footnote{Here, for simplicity, we are considering time-constant thresholds. In later sections, we will show how this notion of action can be used with time-varying thresholds as well.}. This action determines the probability distribution of the reward, $R^a_i(r)$,  and transition time, $D^a_i(d)$, in that trial. The dynamic of the environment is unknown to the participant which means that the participant does not know the relationship between the decision threshold $a$ and the probabilities $R^a_i(r)$ and $D^a_i(d)$. Instead, in each trial, the participant observes reward $r$ and the inter-trial time $d$, which are random samples from these distributions. In the RL model, which we explain later, the participant uses these random samples to adjust her decision threshold after each trial in order to maximize the average reward rate. The two-state SMDP of Experiment A is shown in panel (a) of Figure \ref{fig:SMDP}. 

Finally, we should explain why the average reward rate defined in Equation \ref{eq:Modeling_1} is an appropriate measure of the optimality of the performance in our experiments. An ``optimal participant" in these experiments, will try to obtain the maximum possible reward during the whole experiment. Suppose that a participant sets her decision thresholds in the two conditions such that her overall average reward rate is $\rho$. Since the duration of the experiment is fixed, say $T$ units of time, the participant will receive $\rho \cdot T$ units of reward during the whole experiment. Therefore, to maximize the total amount of reward, the participant should maximize the average reward rate and so this is an appropriate measure of the optimality in our experiments.  

\subsubsection{SMDP model of Experiment B}
As it was explained before, in this experiment no cue is presented at the beginning of the trials. There are two possible ways to model this experiment with an SMDP. In the first model, we assume that since no cue is presented and since the participant sets her decision threshold at the beginning of each trial, all trials are considered by the participant as a single state of the environment. Therefore, the participant will set the same value of the decision threshold for both easy and hard conditions. The corresponding one-state SMDP is shown in panel (b) of Figure \ref{fig:SMDP}.

In the second model, we assume that the participants use a fast mechanism to first detect the difficulty of the trial after the presentation of the stimulus, and then set appropriate decision threshold based on the detected state. In each trial, the environment is in one of two possible states, easy or hard. This true state is unknown (unobservable) to the participant. However, the participant can make observations which can be used to infer, probabilistically, the true state. In Experiment B, these observations are the canoe positions as time elapses in a trial. In the model, it is assumed that in each trial the participant observes the canoe movement for a while and first infers which condition this trial is coming from, and then responds whenever the canoe reaches the threshold corresponding the inferred state. 

For state inference, we use this fact that in the hard trials, the difference between the probability that the canoe moves toward the correct direction and the probability of moving toward the incorrect direction is smaller than in the easy trials. Therefore, the canoe moves back and forth more and so does not get away from the center of the screen as quickly as in the easy trials. Thus, if after some time the canoe is not ``far enough" from the center, it is more likely that the trial is hard. Based on these observations, we propose the following model: the participant observes the canoe movement from the beginning of the trial up to an internal deadline $t_D$. If the absolute value of the canoe position reaches a value $a_D$ at any time $\tau_D$, $0\le \tau_D \le t_D$, the participant infers that the current trial is easy and so sets her decision threshold at values $\pm a_E$. Otherwise, she infers a hard trial and sets the decision threshold at $\pm a_H$. After setting the decision threshold, the participant checks if the canoe position has reached that threshold so far. If this is the case, the participant chooses the right or the left response right away, depending on if the canoe has reached the positive or the negative threshold first. In this case, the decision time will be equal to $\tau_D$. If the canoe position has not reached the decision threshold yet, the participant does not respond and waits until the canoe reaches the decision threshold at some time $t>\tau_D$ and then make her decision. In this case the decision time will be $t$. Therefore, the decision time in each trial is $\max\{t,\tau_D\}$. The values $t_D, a_D, a_E$ and $a_H$ are free parameters of the model and are estimated from each participant's data as we will explain later. The corresponding SMDP is shown in panel (c) of Figure \ref{fig:SMDP}.

\begin{figure}[!h]
	\centering
	\includegraphics[width=.5\linewidth]{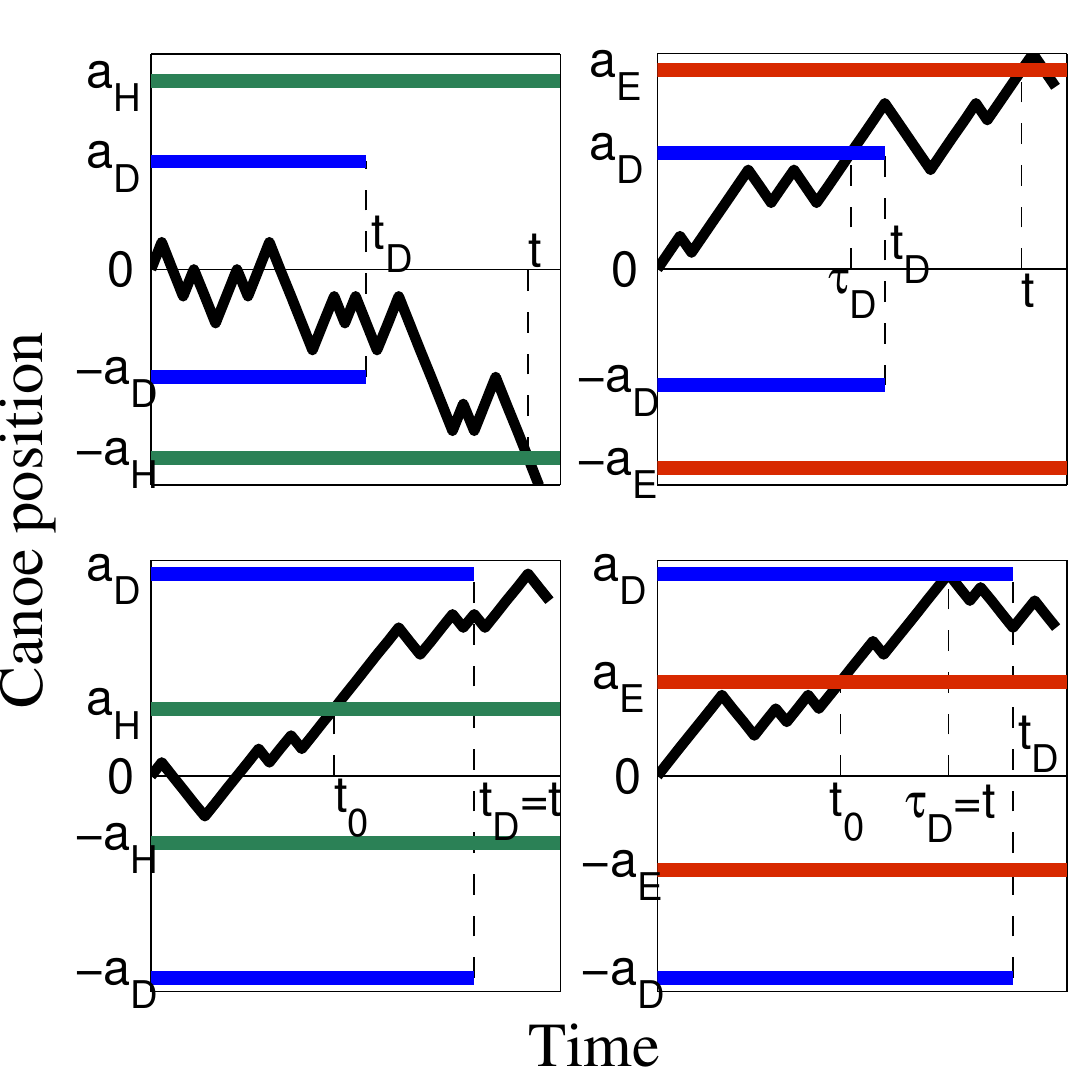}	
	\caption{{\bf Four possible scenarios in detecting the difficulty in Experiment B.} Depending on the values of the thresholds and the canoe path, several cases could happen. In each panel the detected state, the participant's response and the reaction time are as follows: Top-left: $state=H,response=left,RT=t$,Top-right: $state=E,response=right,RT=t$,Bottom-left: $state=H,response=right,RT=t_D$,Bottom-right: $state=E,response=right,RT=\tau_D$. See the text for more details on how to determine these.}
	\label{fig:DiffDetect_B}
\end{figure}

Depending on the values of these parameters and the profile of the canoe position in a trial, several scenarios could happen. Figure \ref{fig:DiffDetect_B} shows four possible scenarios. In the top-left panel of this figure, $a_D<a_H$ and the canoe position did not reach the state detection threshold $a_D$ before the internal deadline $t_D$ and so the participant infers that this is a hard trial and sets her decision threshold at $\pm a_H$. Then, the participant waits until the canoe position reaches $\pm a_H$. In this example, at time $t$ the canoe reaches $-a_H$ and so he participant response is \textit{left} and her decision time is $t$. In the top-right panel, the canoe reaches $a_D$ at $\tau_D$ and so the participant sets her decision threshold at $\pm a_E$. In the scenarios shown in the bottom row of Figure \ref{fig:DiffDetect_B} the sate detection threshold $a_D$ is larger than the decision thresholds. In the bottom-left panel, the canoe does not reach $a_D$ before the deadline $t_D$ and so the threshold is set at $a_H$. However, the canoe has reached $a_H$ at time $t_0<t_D$ and so the decision time will be equal to $t_D= \max(t_0,t_D)$. Similarly, the decision time in the bottom-left panel of the figure is $\tau_D$.
\newline

In both models, the experiment is modeled by a \textit{partially observable} SMDP. In each trial, the environment is in one of two possible states which is unknown to the participant. However, the participant makes noisy observations which can be used to make probabilistic inference about the true state. The models differ in the way these observations are used to make inference and set the decision thresholds. In the second model, as we explained, the participant uses these observation to make a decision about the difficulty of the trial first and then sets the corresponding decision threshold. In the first model, the inference mechanism and its relationship to the shape of the decision threshold is implicit and more complicated. This model has been investigated extensively in previous research \cite{busemeyer_psychological_1988,rao_decision_2010,drugowitsch_cost_2012}. Here, we give a brief explanation of how the single time-varying decision threshold arises as the optimal solution. Suppose that at the beginning of each trial, the participant has some prior belief about the difficulty level of the upcoming trial. When the trial starts, after each time step, the participant receives a piece of noisy information. In the optimal model, the participant uses the Bayes rule after each time step to update her belief about the difficulty of the trial based on the new information. At each time step, the participant must decide if she wants to accumulate information for at least one more time step or she wants to stop accumulating information and respond. The participant makes this decision at each time step based on her current belief about the difficulty of the trial. The participant does not make a decision about the difficulty of the trial. Instead, her belief on how likely each difficulty level is affects her decision at each time step. Intuitively, as time elapses in a trial, it becomes more likely that the trial is hard and so the participant is more willing to stop accumulating information and respond. This results in a time-decreasing decision threshold. In addition, since the observations the participant makes at each time step are considered to be independent, the participants belief at each time step depends only on the value of the accumulated information at that time step and not the history of the observations. Since, the participant's decision at each time step (stop or continue accumulating information) depends only on this belief, this assumption results in a single decision threshold for all trials. In this paper, our focus is not on the mechanism which lead to this single time-varying threshold. Instead, we aimed at investigating if the participants use a single threshold as is suggested by the second model, or they use two thresholds as is suggested by the first model. 

An interesting question that arises is if the maximum possible average reward rate that can be achieved by the second model is equal or smaller than the first model. In the section Optimal performance below, we investigate this question. 

\subsection{Reinforcement learning models of Experiment A}
In this section, we use the SMDP model of experiment A to develop computational models for threshold adjustment in this experiment. In the next section, we propose other computational models which are not based on the SMDP framework.

As it was explained before, the goal of the agent in an SMDP problem is to learn the optimal policy which maximizes the average reward rate. This is equivalent to solving the Bellman optimality equations \ref{eq:Modeling_2}. However, since the dynamic of the environment ($T^a_{i,j}, R^a_i(r)$ and $D^a_i(d)$) is unknown to the agent, it is not possible to solve these equations directly. Instead, the agent should learn the (approximate) optimal policy by interacting with the environment. In the literature of computer science and machine learning, several algorithms based on the theory of reinforcement learning have been proposed to tackle this problem. In one class of these algorithms, the agent uses its experience with the environment to build a model of the environment and uses this model to find the optimal policy. For this reason, these algorithms are known as \textit{model-based RL} algorithms. The main idea is that after each decision (taking an action), the reward, the transition time and the new state given the old state and the taken action, are considered as samples from the corresponding random variables. These samples are used to estimate the probability distributions $T^a_{i,j}, R^a_i(r)$ and $D^a_i(d)$. Therefore, the agent builds an estimate of the dynamic of the environment and updates it after each step. Then, in each state the agent uses this estimate model of the environment and solves Bellman optimality equations using a dynamic programming algorithm. Since the agent has to re-solve the Bellman equations in each step, these algorithms are computationally demanding. On the other hand, since the agent updates its estimate of the dynamic of the environment after each step, model-based RL algorithms are able to find a new optimal policy rather quickly if the environment changes in the middle of the task (for example if the value of the rewards or the transition probabilities changed during the task).

Another class of models, called \textit{model-free RL}, learn the optimal policy by learning the state values directly and without estimating the dynamic of the environment. Comparison between the model-based and model-free algorithms is outside the scope of this paper. Since the action space is continuous in our SMDP models, formulation of the model-based algorithms is harder than the model-free algorithms and so, for the sake of simplicity, we only consider model-free version of our algorithms (see for example \cite{daw_model-based_2011,simon_neural_2011,dezfouli_actions_2013} for more details on comparing model-free and model-based algorithms).  

\begin{figure}[!h]
	\centering
	\includegraphics[width=.8\linewidth]{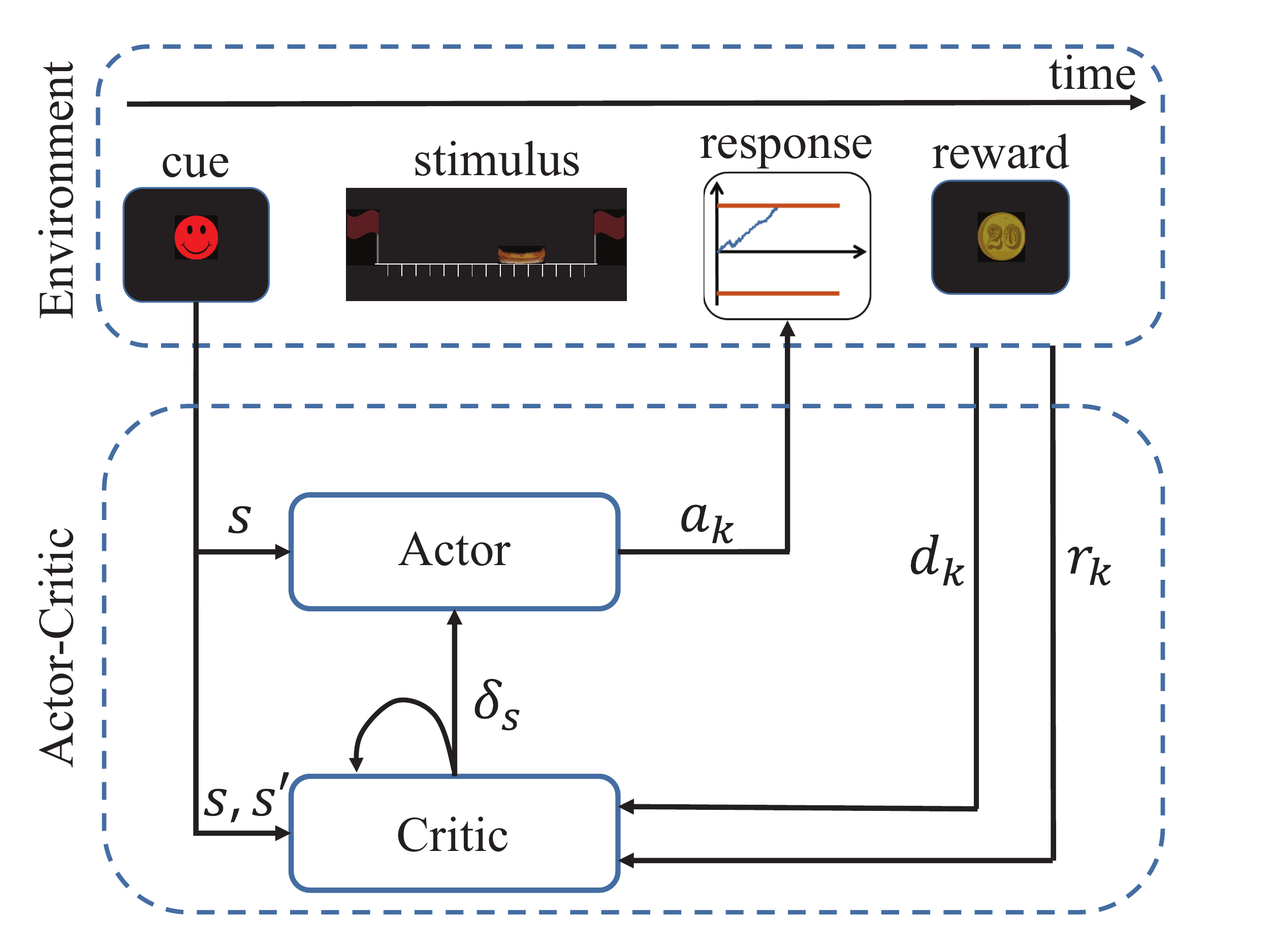}	
	\caption{{\bf Actor-Critic model.} In each trial, the Actor sets the decision thresholds given its current values of the parameters. After responding and receiving the feedback in the trial, the Critic computes the TD error signal. This error is then used to update both the Critic's current estimate of the state values, and the current values of the parameters representing the policy in the Actor.} \label{fig:ActorCriticModel}
\end{figure}

To model the adjustment of the decision thresholds in the SMDP framework, we use an \textit{Actor-Critic} model with \textit{average reward temporal difference learning} (Figure \ref{fig:ActorCriticModel}). In an Actor-Critic architecture, an Actor represents the current policy and is responsible for taking actions in each state, while the Critic module represents the estimated state values and evaluates ``how good" are the actions taken by the Actor based on these estimated values. We first explain the implementation of the Actor module and then the Critic module for our experiments.

In constructing the Actor, we need to consider two problems: how to represent the policy, and how to update the policy. We first explain how the policy is represented in our model and then we turn to the problem of updating policy. 

In most previous psychological experiments for which an RL model has been proposed, in each state there are a few possible actions available to the participant (for example a choice between four doors in \cite{simon_neural_2011} or a choice between two boxes in \cite{daw_model-based_2011}). In these situations, the participant's policy can be specified by a probability mass function $\pi(s,a)=\Pr(a|s), \, \forall a \in A_s$, where $A_s$ is the set of all possible actions in state $s$. In each state $s$, the Actor selects an action $a$ with probability $\pi(s,a)$. After each step, the agent updates the value of these probabilities by some learning rule. In our experiments, the available actions in each state are all possible values for the decision threshold in that state, and therefore, the action space is continuous. Therefore, it is not possible to update the policy as in the discrete action case. One possible way to address this problem is to represent the policy as a parametric distribution over the actions and update the parameters of that distribution \cite{williams_simple_1992}. We first explain the method for the case where the threshold is constant during a trial and then extend it to the more complicated case where the threshold is allowed to dynamically change within a trial (i.e. time-varying threshold).

The simplest choice is to represent the policy by a Gaussian distribution:

\begin{equation}
\pi_k(s,a)=\frac{1}{\sqrt{2\pi \sigma_{k,s}^2}}\exp\bigg( \frac{(a-m_{k,s})^2}{2\sigma_{k,s}^2} \bigg)
\label{eq:RL_models_01}
\end{equation}    

This means that if trial $k$ is from condition $s$ (easy or hard), the participant's threshold will be a Gaussian random variable with mean $m_{k,s}$ and variance $\sigma_{k,s}^2$. The actual value of the decision threshold, that we observe, in that trial will be a sample from this distribution. The participant updates the mean and variance of this distribution after each trial by some learning rule. Even if there is no learning, that is if $m_{k,s}$ and $\sigma_{k,s}^2$ remain constant during the experiment, representing the policy by a random variable means that there is between-trial variability in the decision threshold. Before explaining the learning rule for the policy parameters, it is important to justify the choice of probabilistic decision thresholds in our models. In most instantiations of the sequential sampling models, the decision threshold is assumed to be a deterministic parameter with no between-trial variability. These models, however, assume that some other parameters are random variables (for example initial value of the accumulated information and the drift rate in the drift diffusion mode (DDM)). The main reason for this modeling assumption is that these models are used for experiments in which the decision threshold is not observed directly (e.g., the random dot motion task) and the effect of variability in the decision threshold can be mimicked by variability in other parameters (specifically with variability in the initial value of the accumulated information in DDM). Therefore, these models arbitrarily assume that the decision threshold is deterministic and some other parameters are random\footnote{One notable exception in this regard is the Grice modeling framework in which the decision process is modeled as a race between several accumulators \cite{grice_stimulus_1968}. In this framework, the information accumulation in each accumulator is modeled as a deterministic function while the thresholds are modeled as random variables (see \cite{jones_unfalsifiability_2014} for the relationship between this framework and some other sequential sampling models).}. In our experiments, in contrast, we can directly observe the decision threshold and, as we will see, there is a lot of between-trial variability in the decision threshold of each participant. In addition, in this experiment, the initial position of the canoe is zero in all trials and so there is no variability in the initial value of the accumulated information. Finally, the rate of the information accumulation is controlled by the probability of moving in the correct direction in each time step, and so it remains constant between trials. For these reasons, representing the decision threshold as a random variable is reasonable in our models.

Now we turn to the problem of updating the policy parameters after each trial. Developing algorithms for learning the optimal policy for continuous action-space problems is challenging, specifically when the dimensionality of the state-space increases. In the SMDPs of our experiments, however, there are at most 2 states. Also, our goal is not to develop an algorithm that can efficiently learn the optimal policy. Instead, we are interested in simple algorithms that can model human participants' learning mechanisms in our experiments. Here, we develop a simple instantiation of a class of algorithms called \textit{REINFORCE} \cite{williams_simple_1992}. Consider a stochastic policy $\pi(s,a|\theta_s)$, in which the possible actions in state $s$ are parameterized by the set of parameters $\theta_s=[\theta_{1,s},...,\theta_{p,s}]^T$. In a \textit{REINFORCE} learning algorithm, the amount of change in each parameter after transition from state $s$ in trial $k$, to a new state $s'$ is:

\begin{equation}
\Delta\theta_{i,s,k}=\alpha_{s,k} \cdot (r_{s}-B_{i,s}) \cdot  \frac{\partial \ln[\pi(s,a|\theta_{s,k})]}{\partial \theta_{i,s,k}}
\label{eq:RL_models_02}
\end{equation}
 
 \noindent where $\alpha_{s,k}$ is the learning rate, $r_{s}$ is some measure of the future reward achievable from state $s$ by following the current policy, and $B_{i,s}$ is a \textit{baseline reinforcement}. Williams \cite{williams_simple_1992} proved that in any \textit{REINFORCE} algorithm: $E[\Delta \theta_s|\theta_s]=\alpha_{s} \cdot \nabla_{\theta_s} E[r_s|\theta_s]$. In words, moving in the direction of $\Delta \theta_s$, on average, is equivalent to moving in the direction of the gradient of the average future reward with respect to the parameters.
 
The learning rule \ref{eq:RL_models_02} defines a general class of algorithms. To specify a learning algorithm in this class completely, one needs to define the measure of the future reward $r_s$, and the baseline reinforcement $B_{i,s}$. In the SMDP framework, the future expected reward is specified by the state values, $V(s)=E[r_{s,s'}-\rho\cdot\tau+V(s')]$, where $r_{s,s'}$ is the one-step reward that is achieved by going from $s$ to $s'$. The expectation is taken over the policy and all possible future states $s'$ and dwell times $\tau$. Computing this expectation needs the knowledge of the quantities $T^a_{i,j}, R^a_i(r)$ and $D^a_i(d)$ (see Equation \ref{eq:Modeling_2}). Instead, we use the one-step sample of this expectation. Suppose that at trial $k$ the agent is in state $s$, takes action $a_k$, and transitions to a new state $s'$ after $d_k$ units of time. Also, suppose that the agent's current estimate of the value of states $s$ and $s'$, and the average reward rate are $\hat{V}_k(s),\, \hat{V}_k(s')$ and $\hat{\rho}_k$, respectively. Then, the estimate of the expected future reward in state $s$ is defined as $r_{s,s'}-\hat{\rho}\cdot d_k + \hat{V}_k(s')$. If in addition, we defined $B_{i,s}=\hat{V}_k(s)$, the term in the parentheses in Equation \ref{eq:RL_models_02} becomes:

\begin{equation}
\delta_s(k)=r_{s,s'}-\hat{\rho}_k\cdot d_k + \hat{V}_k(s')-\hat{V}_k(s)
\label{eq:RL_models_09}
\end{equation}

$\delta_s(k)$ is called the \textit{temporal difference (TD) error} and plays an important role in many reinforcement learning algorithms. By computing the derivative of the natural logarithm of the right-hand sight of Equation \ref{eq:RL_models_01} with respect to $m_{k,s}$ and replacing it in Equation \ref{eq:RL_models_02}, the final form of the learning rule for $m_{k,s}$ will be:

\begin{equation}
m_{k+1,s}=m_{k,s} + \alpha_{s,m} \cdot \delta_s(k) \cdot  (a_k-m_{k,s})
\label{eq:RL_models_03}
\end{equation}     
 
 \noindent where $a_k$ is the observed decision threshold in trial $k$. In this equation, as it was suggested by \cite{williams_simple_1992}, we have set $\alpha_{s,k}=\alpha_{s,m}\cdot \sigma_{s,k}^2$. It is easy to derive a similar learning rule for $\sigma_{s,k}^2$. However, to simplify the model, we assume that this parameter increases or decreases linearly during the experiment:
 
 \begin{equation}
\sigma_{k,s}^2=\sigma_{0,s}^2 + \alpha_{s,\sigma} \cdot k
 \label{eq:RL_models_04}
 \end{equation} 
 
Here, $\sigma_{0,s}^2$ and $\alpha_{s,\sigma} \in [-1,1]$ are free parameters of the model. The value of $\sigma_{s,k}^2$ determines the balance between \textit{exploration}  and \textit{exploitation} in the model: larger values of this parameter means that the participant chooses values different from the mean more often and therefore explores the policy space more. On the other hand, smaller values of this parameter means that the participant uses values closer to the mean and therefore exploits what she has learned instead of exploration. We expect that the participants explore less with experience and so the value of $\sigma_{s,k}^2$ decreases. However, some participants may decide to explore more with more experience and therefore we allow the parameter $ \alpha_{s,\sigma}$ to take both positive and negative values. 

In the Actor-Critic architecture, the TD error signal is computed by the Critic. This module keeps the current estimates of the state values and updates them using the following \textit{TD learning} rule:

 \begin{equation}
\hat{V}_{k+1}(s)=\hat{V}_{k}(s)+\alpha_{s,c} \cdot \delta_s(k)
 \label{eq:RL_models_05}
 \end{equation} 

The architecture of the model is depicted in Figure \ref{fig:ActorCriticModel}. This figure shows the interaction between the agent, modeled as the Actor-Critic architecture, and the environment. After the presentation of the cue at the beginning of a trial the state $s$ is determined and both the Actor and the Critic are informed about it. The Actor then choses a value for the action based on its current policy. Specifically, it draws a sample from the Gaussian distribution \ref{eq:RL_models_01}, given the current values of the parameters $m_{s,k}$ and $\sigma_{s,k}^2$. This value, $a_k$, is set as the decision threshold for the current trial. The participant responses as soon as the position of the canoe exceeds this value on either side of the screen. Based on the chosen response and the state, the participant receives some reward $r_k$. The amount of this reward and the total time of the trial, $d_k$, are sent to the Critic. After the presentation of the cue in the next trial, the Critic uses its current estimate of the values of the previous and current states, the reward and trial time in the previous state, and its current estimate of the average reward rate, to compute the TD error. This signal, then, is used to update both the state values, using Equation \ref{eq:RL_models_05}, and the policy parameters, using Equations \ref{eq:RL_models_03} and \ref{eq:RL_models_04}.\newline  

So far, we have assumed that the decision thresholds remain constant within a trial. Next, we extend this model to the case where the decision thresholds can change dynamically within a trial. We can still represent the policy with the Gaussian distribution \ref{eq:RL_models_01}, but now $m_{s,k}$ is a function of the elapsed time in the trial, and so we denote it by $m_{s,k}(t)$. We need to represent $m_{s,k}(t)$ with a parametric function which is flexible enough to generate different patterns that the decision threshold may take, and has few parameters so the number of the free parameters of the model stays small. Following \cite{hawkins_revisiting_2015} we use the following functional form for the thresholds which is the Weibull cumulative distribution function:

\begin{equation}
m_{s,k}(t)=\psi_{s,k}-\bigg[1-\exp\bigg(-\bigg(\frac{t}{\lambda_{s}} \bigg)^{\phi_{s}} \bigg) \bigg]\cdot \bigg[\frac{1}{2} \psi_{s,k} - \psi_{s}' \bigg]
\label{eq:RL_models_06}
\end{equation} 

The free parameters are: $\psi_{s,k}, \, \lambda_{s}, \, \phi_{s}$ and $\psi_{s}'$. To keep the number of free parameters low, we assume that the participant only learns $\psi_{s,k}$, and other parameters of this function remain constant during the experiment. To specify the learning rule for $\psi_{s,k}$, we first need to compute $\frac{\partial \ln[\pi(s,a|\theta_{s,k})]}{\partial \psi_{s,k}}$ . We have:

\begin{equation}
\frac{\partial \ln[\pi(s,a|\theta_{s,k})]}{\partial \psi_{s,k}}=\frac{\partial \ln[\pi(s,a|\theta_{s,k})]}{\partial m_{s,k}}\cdot \frac{\partial m_{s,k}}{\partial \psi_{s,k}}=\frac{(a_k-m_{s,k})}{\sigma_{s,k}^2} \cdot \bigg[0.5+0.5\exp\bigg(-\bigg(\frac{t_k}{\lambda_{s}} \bigg)^{\phi_{s}} \bigg) \bigg]
\label{eq:RL_models_07}
\end{equation} 

\noindent where $t_k$ is the time to complete trial $k$. Therefore, we use the following learning rule:

\begin{equation}
\psi_{s,k+1}=\psi_{s,k}+\alpha_{s,\phi} \cdot \delta_s(k) \cdot (a_k-m_{s,k}) \cdot \bigg[1+\exp\bigg(-\bigg(\frac{t_k}{\lambda_{s}} \bigg)^{\phi_{s}} \bigg) \bigg]
\label{eq:RL_models_08}
\end{equation} 

\noindent We also use learning rule \ref{eq:RL_models_04} for $\sigma_{s,k}^2$.

\begin{figure}[!h]
	\centering
	\includegraphics[width=.8\linewidth]{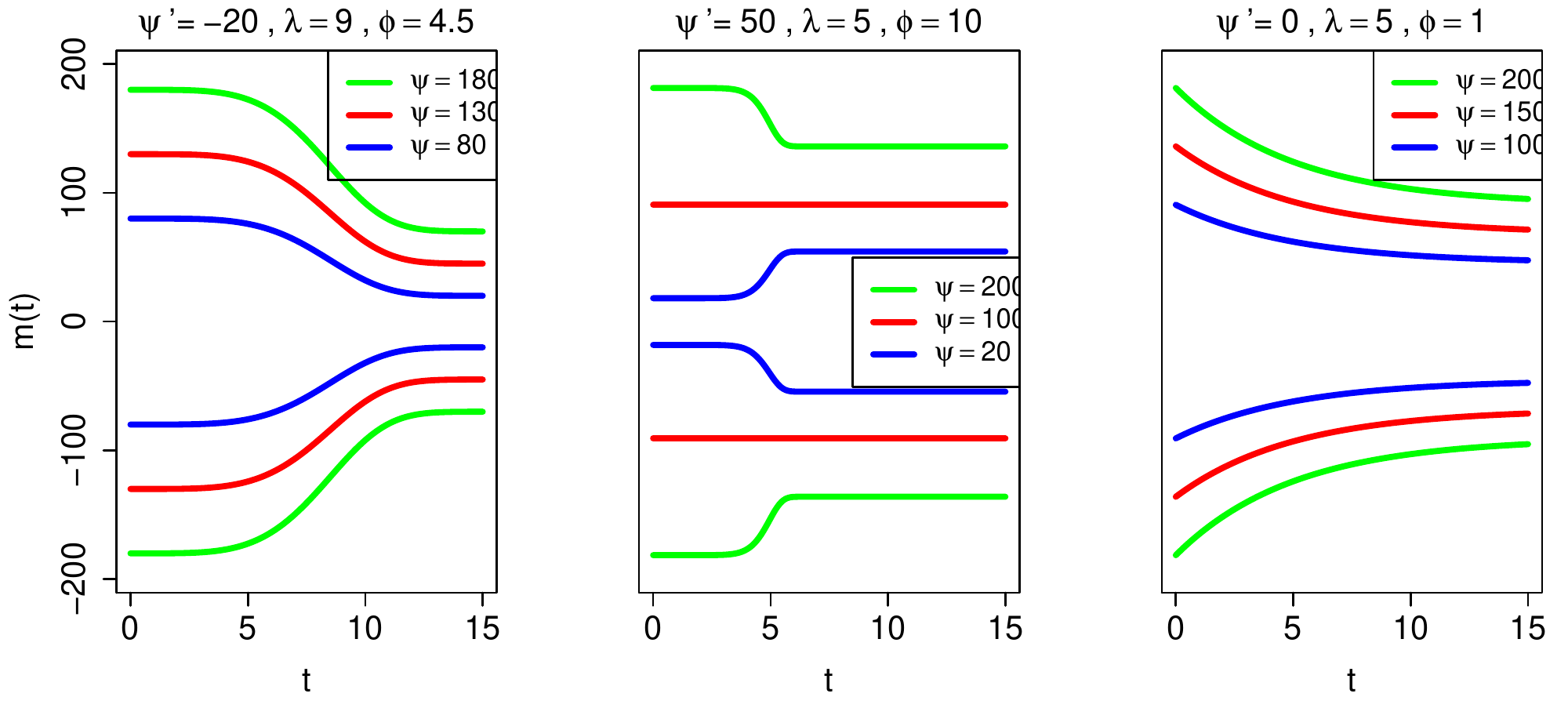}	
	\caption{{\bf Time-varying decision thresholds.} The shape of Weibull function for different values of the parameters. In each panel $\psi',\,\lambda$ and $\phi$ are kept constant and the function is plotted for 3 values of $\psi$. As it can be seen, changing $\psi$ may change the shape of the function.} \label{fig:WeibullThreshold}
\end{figure}

In Figure \ref{fig:WeibullThreshold}, the function in Equation \ref{eq:RL_models_06} is plotted for different values of its parameters. Two points should be noted in this figure: First, for different values of the parameters, this function can take different forms. Second, by changing $\psi$ the shape of the function may change dramatically. For example, in the middle panel of this figure, for $\psi=200$ the function is decreasing, for $\psi=100$ it is constant, and for $\psi=20$ it is increasing (other parameters are kept constant). 

Finally, we assume that the subjective value of reward is proportional to the number of coins obtained in each trial. Specifically, we let $r_k=\beta_{s,r}\cdot x_k$, where $r_k$ is the subjective reward, $x_k$ is the number of coins, and $\beta_{s,r}$ is the coefficient for condition $s=E$ or $H$. The effect of this transformation, however, can be captured by $\hat{\rho}$, $\alpha_{s,c}$ and $\alpha_{s,\phi}$ and so we do not consider $\beta_{s,r}$ as free parameters. The model with time-constant thresholds has 13 free parameters, $(\sigma_{s,0}^2,m_{s,0},\alpha_{s,\sigma},\alpha_{s,m},\alpha_{s,c},\hat{V}_{s,0},\hat{\rho})$,while the model with time-varying thresholds has 19 free parameters, $(\sigma_{s,0}^2,\psi_{s,0},\psi_{s}',\lambda_{s},\phi_{s},\alpha_{s,\sigma},\alpha_{s,c},\alpha_{s,\phi},\hat{V}_{s,0},\hat{\rho})$\footnote{In the RL models of both experiments A and B we assume that $\hat{\rho}$ is a fixed value and estimate it as a free parameter for each participant. This is not a plausible assumption. The value of the reward rate should be estimated after each trial using the values of the rewards and trial times. We fitted a version of the RL models in which $\hat{\rho}$ was estimated using a simple moving average algorithm. However, this model resulted in larger BIC compared to the simpler RL models. Therefore, we do not report the results of these models here. One reason that models with constant $\hat{\rho}$ resulted in lower BIC could be that in the experiments the value of the reward rate remains constant and therefore the additional complexity in the model for estimating $\hat{\rho}$ is not necessary.}.

\subsection{Other computational models of Experiment A}
The models developed in the previous section can, potentially, learn the optimal policy that leads to the maximum possible average reward rate. Although computationally appealing, they may not be able to explain human participants' learning behavior in our experiments. Also, not all participants use the same learning strategy. Some participants may set their decision threshold at the beginning of the experiment and use the same threshold for the rest of the experiment. Also, in the RL models of the previous section, it is assumed that the participants consider a cost for time. This is in contrast to the models which have been proposed to explain the \textit{post-error slowing effect} in perceptual decision making tasks. Based on these models, the participants increase (decrease) their decision threshold after an incorrect (correct) response \cite{laming_choice_1979,smith_slowness_1995} and only based on the values of the rewards. Finally, the participants may set two different desired levels of accuracy and reaction time for the easy and hard conditions, based on the instructions at the beginning of the experiment, and adjust their decision threshold to achieve those desired levels. In this section, we will develop models for Experiment A based on these assumptions. 

In all these models, the decision threshold is modeled as a Gaussian random variable with the p.d.f. of Equation \ref{eq:RL_models_01}. In the baseline models below, the mean $m_{s,k}$ and variance $\sigma_{s,k}^2$ remain constant during the experiment. For the other models, $\sigma_{s,k}^2$ is updated using Equation \ref{eq:RL_models_04}. Therefore, the only difference between these models and the RL models in the previous section is in the way $m_{s,k}$ is updated.

\subsubsection{Baseline models}
The simplest possible model for the experiment is to assume that no learning occurs during the experiment. We consider two versions of this model. In the first version, which we call Model 0 for future reference, the participant sets two different time-constant decision threshold for the two conditions. This model has 4 free parameters, $(m_E,\sigma_E^2,m_H,\sigma_H^2)$, which are the mean and variance of the decision thresholds for the easy and hard conditions.

In the second version, Model 1, the participants sets two time-varying decision thresholds for the two conditions. The decision thresholds have the functional form of Equation \ref{eq:RL_models_06}. This model has 10 free parameters: $(\sigma_s^2,\psi_s,\psi_{s}',\lambda_{s},\phi_{s})$, for $s=E,H$.

\subsubsection{Adjusting thresholds proportional to reward}
We will consider three versions of this model. In all these versions, the threshold is adjusted as a function of the subjective value of the reward received in a trial. Specifically, these models assume that if the participant receives a positive reward in a trial, she will reduce her decision threshold for the next trial, while she will increase her decision threshold after receiving a negative reward. Let $x_k$ be the number of coins that the participant receives or loses in trial $k$. This will cause a subjective value of reward $r_k$ where:

\begin{equation}
\label{eq:A_model_reward_1}
r_k=\left\{\begin{matrix}
[1-exp(-\beta_p \cdot x_k)] / \beta_p, & x_k \ge 0\\ 
[1-exp(-\beta_n \cdot x_k)] / \beta_n, & x_k < 0
\end{matrix}\right.
\end{equation}

The parameters $\beta_p$ and $\beta_n$ control the shape of the subjective reward function (also called the \textit{utility function}) for the positive and negative rewards, respectively.

In the first version of the model, Model 2, the participant uses two different time-constant decision thresholds for the easy and hard trials and adjusts them independently. At the beginning of the experiment, the mean of these decision thresholds are $m^0_s$, $s=E$ or $H$. After each subsequent trial $k$ of condition $s$, the mean of the corresponding threshold is updated as follows:

\begin{equation}
\label{eq:A_model_reward_2}
m_s^{k+1}=m_s^k + \alpha_s \cdot r_k
\end{equation}

\noindent where $\alpha_s>0$ is the learning rate in condition $s$, and $r_k$ was defined in Equation \ref{eq:A_model_reward_1}. This model has 10 free parameters, $(m_{s,0},\sigma_{s,0}^2,\alpha_{s,\sigma},\alpha_s,\beta_p,\beta_n)$ for $s=E,H$.

In the second version of the model, Model 3, the participant uses time-varying thresholds. In contrast to the RL models, we do not assume that the participant learns the parameters of these functions. Instead, the whole function is increased or decreased after each trial:

\begin{equation}
\label{eq:A_model_reward_3}
m_s^{k+1}(t)=m_s^k(t) + \alpha_s \cdot r_k
\end{equation}

\noindent This model has 16 free parameters: $(\sigma_{s,0}^2,\alpha_{s,\sigma},\psi_s,\psi_{s}',\lambda_{s},\phi_{s},\alpha_s,\beta_p,\beta_n)$

In models 2 and 3, the decision thresholds in the two conditions are adjusted independently. That is, if for example trial $k$ is an easy trial, the threshold for the hard condition will not be updated after this trial. In the third version of this class of models, Model 4, we assume that the reward received in one condition affects the threshold in the other condition. Specifically, if trial $k$ is from condition $s$, the thresholds corresponding to this condition and condition $s'\ne s$ are updated as follows:

\begin{equation}
\label{eq:A_model_reward_4}
\begin{matrix}
m_{s,k+1}(t)=m_{s,k}(t) + \alpha_s \cdot r_k \\
m_{s',k+1}(t)=m_{s',k}(t) + \gamma_{s'} \cdot r_k
\end{matrix}
\end{equation} 

\noindent When $\gamma_s=0$ this model reduces to Model 3. It is worth noting that in the RL models, since the TD error $\delta_s(t)$ is a function of both $V(s)$ and $V(s')$, the rewards in one condition affect the adjustment of the threshold in the other condition.

\subsubsection{Adjusting thresholds to achieve desired level of accuracy}
The participant's objective in this model is different from the previous models. In this model, it is assumed that the participant has a desired level of accuracy for each condition. After each trial, the participant lowers her decision threshold if her current estimate of accuracy is higher than the desired level and increases her threshold otherwise. Mathematically, after trial $k$ the thresholds are updated as follows:

\begin{equation}
\label{eq:A_model_acc_1}
m_{s,k+1}(t)=m_{s,k}(t) + \alpha_s \cdot (\theta_s-\hat{p}_{s,k})
\end{equation}
 
 In this equation, $\alpha_s$ is the learning rate, $\theta_s$ is the desired level of accuracy, and $\hat{p}_{s,k}$ is the participant's estimate at trial $k$ of her accuracy in condition $s$. 

We assume that the participant has a prior belief about her accuracy in each condition, and updates this belief after each trial using the Bayes rule. The prior belief is modeled as a Beta distribution, $Beta(a_s,b_s)$. Assume that up to trial $k$, the participant has experienced condition $s$, $n_s$ times of which $n_{s,c}$ times her responses have been correct (and $n_s-n_{s,c}$ incorrect). Then, the posterior belief after trial $k$ will be a Beta distribution, $Beta(a_s+n_{s,c},b_s+n_s-n_{s,c})$. We assume that the participant uses the mean of this posterior as her estimate of the accuracy and so:

\begin{equation}
\label{eeq:A_model_acc_2}
\hat{p}_{s,k}=\frac{a_s+n_{s,c}}{a_s+b_s+n_s}
\end{equation}  

The version of the model with time-constant thresholds, Model 5, has 14 free parameters, $(m_{s,0},\sigma_{s,0}^2,\alpha_{s,\sigma},\alpha_s,a_s,b_s,\theta_s)$, and the version with time-varying thresholds, Model 6, has 20 free parameters, $(\sigma_{s,0}^2,\psi_s,\psi_{s}',\lambda_{s},\phi_{s},\alpha_{s,\sigma},\alpha_s,a_s,b_s,\theta_s)$.

\subsubsection{Adjusting thresholds to achieve desired level of accuracy and mean RT}
This model can be considered as a generalization of Model 6. The participant's objective in this model is to achieve a desired level of both accuracy and mean RT. Let $\theta_s$ and $T_s$ denote the desired levels of accuracy and mean RT for condition $s$, respectively. After each trial $k$, the participant has an estimate of the current accuracy and mean RT in each condition. To keep the number of the free parameters of the model low, we assume that the estimate of the accuracy is obtained by taking the average of the number of correct trials in the last $w_a$ trials, and the estimate of the mean RT is obtained by taking the average of the RT in the last $w_t$ trials, where $w_a$ and $w_t$ are free parameters. Let $\hat{p}_{s,k}$ and $\hat{t}_{s,k}$ be the estimates of the accuracy and mean RT for condition $s$ after trial $k$. These estimates are obtained as follows:

\begin{equation}
\label{eeq:A_model_acc_t_1}
\hat{p}_{s,k}=\frac{n_{c,s}}{w_a}
\end{equation} 

\begin{equation}
\label{eeq:A_model_acc_t_2}
\hat{t}_{s,k}=\frac{\sum_{i=k-w_t+1}^{k}d_i}{w_a}
\end{equation} 

\noindent where $n_{c,s}$ is the number of correct responses in the last $w_a$ trials, and $d_i$ is the total time spent on trial $i$.

Now the question is how the participants must adjust their decision thresholds to achieve the desired levels of accuracy and mean RT. We assume that the participants know that by increasing the decision threshold both their accuracy and mean RT increases and vice versa (this information is given to the participants explicitly in the instructions). Now suppose that both the current estimates of the accuracy and the mean RT are lower than their corresponding desired level. In this case, to become closer to the desired levels, the participant must increase her decision threshold. But what if, for example, the current estimate of the accuracy is lower than the desired level while the current estimate of the mean RT is higher than the desired level? 

Here, we consider a simple rule for updating the mean of the decision threshold:

\begin{equation}
\label{eq:A_model_acc_t_3}
m_{s,k+1}(t)=m_{s,k}(t) +\Delta m_{s,k}
\end{equation}

\noindent where:

\begin{equation}
\label{eq:A_model_acc_t_4}
\Delta m_{s,k}=\alpha_a\cdot (\hat{p}_{s,k}-\theta_s)+\alpha_t\cdot (\hat{t}_{s,k}-T_s)
\end{equation}

In words, the change in the threshold is the weighted sum of the discrepancies between the current estimates and the desired levels of the accuracy and mean RT. The weights $\alpha_a$ and $\alpha_t$ can be considered as ``attentional weights": the relative amount of attention that the participant pays to the discrepancy in the ``dimensions" accuracy and RT. The notion of ``attentional weights" is used widely in the computational models of categorization \cite{nosofsky_attention_1986}. 

Not all free parameters in equation \ref{eq:A_model_acc_t_4} are identifiable. Therefore, we use the following equation for model fitting which is obtained by simple algebraic manipulation of equation \ref{eq:A_model_acc_t_4}:

\begin{equation}
\label{eq:A_model_acc_t_5}
\Delta m_{s,k}=\alpha_a\cdot \hat{p}_{s,k}+\alpha_t\cdot \hat{t}_{s,k}+c_{s}
\end{equation}

We only consider a version of this model with time-varying thresholds. This model, which we call Model 7, has 18 free parameters, $(\sigma_{s,0}^2,\psi_s,\psi_{s}',\lambda_{s},\phi_{s},\alpha_{s,\sigma},\alpha_t,\alpha_a,c_s,w_t,w_a)$ .

It is important to note that this model does not provide a normative account of the decision making process. For example, if a participant has a very high level of desired accuracy and pays too much attention to this dimension (large value of $\alpha_a$) she will eventually set her decision thresholds at values higher than the optimal values. Model 7 is meant to provide a reasonable ``descriptive" account of the learning process. This provides an important alternative to the RL models which are based on the optimization of performance. 

It is not hard to see that if $w_a=w_t=1$ and if there was only one state in the experiment, then Model 7 and the RL model described above will be equivalent. Therefore, to be able to compare these models it is important to have two types of trials and the cue at the beginning of the trials which creates an SMDP with two states as we saw before. Computational models for Experiment A are summarized in Table \ref{tab:models_A}. In this table, the RL models with time-constant and time-varying thresholds are called Model 8 (or $RL_C$) and 9 (or $RL_V$), respectively.

\begin{table}[h]
	\centering
	\caption{Computational models of Experiment A}
	\label{tab:models_A}
	
	\begin{center}
		\begin{threeparttable}
			\begin{tabular}{lccc}
				\toprule		
				\textbf{Model} & \textbf{Threshold\tnote{1}} & \textbf{Parameters} & \parbox[t]{2cm}{\textbf{No. of free\\parameters}} \\ \midrule
				0 & C & $\sigma_s^2,m_s$ & 4\\[1pt]
				1 & V & $\sigma_s^2,\psi_s,\psi_{s}',\lambda_{s},\phi_{s}$ & 10 \\ \midrule
				2 & C & $\sigma_{s,0}^2,m_{s,0},\alpha_{s,\sigma},\alpha_s,\beta_p,\beta_n$ & 10 \\[1pt]
				3 & V &  $\sigma_{s,0}^2,\alpha_{s,\sigma},\psi_s,\psi_{s}',\lambda_{s},\phi_{s},\alpha_s,\beta_p,\beta_n$ & 16 \\[1pt]
				4 & V &  $\sigma_{s,0}^2,\alpha_{s,\sigma},\psi_s,\psi_{s}',\lambda_{s},\phi_{s},\alpha_s,\beta_p,\beta_n,\gamma_s$ & 18  \\ \midrule		
				5 & C & $\sigma_{s,0}^2,m_{s,0},\alpha_{s,\sigma},\alpha_s,a_s,b_s,\theta_s$ & 14  \\[1pt]
				6 & V &  $\sigma_{s,0}^2,\psi_s,\psi_{s}',\lambda_{s},\phi_{s},\alpha_{s,\sigma},\alpha_s,a_s,b_s,\theta_s$ & 20\\ \midrule
				7 & V &  $\sigma_{s,0}^2,\psi_s,\psi_{s}',\lambda_{s},\phi_{s},\alpha_{s,\sigma},\alpha_t,\alpha_a,c_s,w_t,w_a$ & 18\\ \midrule
				$\mathrm{8 \, (RL_C)}$ & C & $\sigma_{s,0}^2,m_{s,0},\alpha_{s,\sigma},\alpha_{s,m},\alpha_{s,c},\hat{V}_{s,0},\hat{\rho}$ & 13 \\
				$\mathrm{9 \, (RL_V)}$ & V &  $\sigma_{s,0}^2,\psi_{s,0},\psi_{s}',\lambda_{s},\phi_{s},\alpha_{s,\sigma},\alpha_{s,c},\alpha_{s,\phi},\hat{V}_{s,0},\hat{\rho}$ & 19  \\ \bottomrule					
			\end{tabular}
			\begin{tablenotes}
				\item[1] C:time-constant, V:time-varying
			\end{tablenotes}
		\end{threeparttable}
	\end{center}
\end{table}

\subsection{Computational models of Experiment B}
The goal here is to assess two hypotheses regarding the decisional mechanism of the participants. Based on the first hypothesis, $H_1$, since no cue is presented at the beginning of the trials, the participants use the same value of the decision threshold for both easy and hard trials. As it was explained before, the corresponding SMDP model in this case will have only one state. This hypothesis is particularly interesting due to the fact that the optimal behavior in this experiment is to use one set of time-decreasing decision threshold for both types of trials \cite{drugowitsch_cost_2012,rao_decision_2010}. We will show this in Section Optimal performance below.

Based on the second hypothesis, $H_2$, the participants use a \textit{difficulty-detection} mechanism at the beginning of the trial to identify the difficulty of the current trial and then set their decision threshold accordingly.

To compare these hypotheses, we consider three models based on the first hypothesis and one model based on the second hypothesis. In the first model based on $H_1$, $M_{H_1}^1$,the participants use a time-varying decision threshold, which is implemented by the functional form \ref{eq:RL_models_06}, for both conditions. In addition, we do not consider any learning in this model. This model has 5 free parameters: $(\sigma^2,\psi,\psi',\lambda,\phi)$.

In the second model for $H_1$, $M_{H_1}^2$, the threshold is modeled the same way. The only difference here is that in this model we use a one-state version of the RL model to adjust the mean of the thresholds, $m_{s,k}(t)$. When there is only one state, $s=s'$ in Equation \ref{eq:RL_models_09} and so the TD error will be $\delta_k=r_k-\hat{\rho}\cdot d_k$, and does not depend on the state value anymore. We use this TD error together with Equation \ref{eq:RL_models_03} to update the mean of the Gaussian distributions after each trial. We also assume that the subjective value of positive and negative rewards are $\beta_p$  and $\beta_n$, respectively. This model has 9 free parameters:  $(\sigma^2,\psi,\psi',\lambda,\phi,\alpha_{\sigma},\beta_p,\beta_n,\hat{\rho})$.

In the third model based on $H_1$, $M_{H_1}^3$, as in the previous two models, the participants use the same threshold for both types of trials. However, in contrast to $M_{H_1}^2$, here the experiment is modeled as a two-state SMDP (this SMDP is not shown in Figure \ref{fig:SMDP}). The main assumption is that the participants sets their decision threshold at the beginning of each trial and so they have to use the same decision threshold for both types of trials. However, the adjustment of this single threshold is carried out after the trial is finished and so the participant knows what was the trial condition. In other words, the experiment is modeled as a two-state SMDP but the available action is the same in both states. This means that, in contrast to  $M_{H_1}^2$,  $\hat{V}_k(s')$ and $\hat{V}_k(s)$ are not canceled out and the TD error will be computed using Equation \ref{eq:RL_models_09}. This model has 12 free parameters: $\sigma^2,\psi,\psi',\lambda,\phi,\alpha_{s,m},\alpha_{s,m},\hat{V}_{s,0},\hat{\rho}$.   

In the model based on $H_2$, $M_{H_2}$, the participant first observed the canoe movement up to an internal deadline $t_D$. If before this deadline the canoe position exceeds a difficulty detection threshold $a_D$, the participant decides that the current trial is easy and sets her decision thresholds at $\pm a_E$. Otherwise the trial is detected as being hard and the participant uses the thresholds $\pm a_H$. As before, the decision thresholds are modeled as Gaussian random variables\footnote{To reduce the number of free parameters, we assume that $a_D$ is a deterministic variable.}. We only consider a version of this model in which the thresholds $a_E$ and $a_H$ are time-constant. Specifically, if trial $k$ is detected to be from the condition $s$, the value of the decision threshold is a sample from a Gaussian variable with mean $m_{s,k}$ and variance $\sigma_s^2$. The value of the means are updated after each trial using Equations \ref{eq:RL_models_03} and \ref{eq:RL_models_05}. This model has 13 free parameters: $(\sigma_s^2,m_{s,0},\alpha_{s,c},\alpha_{s,m},\hat{V}_{s,0},\hat{\rho},t_D,a_D)$. Computational models of Experiment B are summarized in Table \ref{tab:models_B}.

\begin{table}[h]
	\centering
	\caption{Computational models of Experiment B}
	\label{tab:models_B}
	
	\begin{center}
		\begin{threeparttable}
			\begin{tabular}{lcc}
				\toprule		
				\textbf{Model} & \textbf{Parameters} & \parbox[t]{2cm}{\textbf{No. of free\\parameters}} \\ \midrule
				$M_{H_1}^1$  & $\sigma^2,\psi,\psi',\lambda,\phi$ & 5\\[1pt]
				$M_{H_1}^2$ & $\sigma^2,\psi,\psi',\lambda,\phi,\alpha_{\sigma},\beta_p,\beta_n,\hat{\rho}$ & 9 \\ [1pt]
				$M_{H_1}^3$ & $\sigma^2,\psi,\psi',\lambda,\phi,\alpha_{s,m},\alpha_{s,m},\hat{V}_{s,0},\hat{\rho}$ & 12 \\ [1pt] \midrule
				$M_{H_2}$ &  $\sigma_s^2,m_{s,0},\alpha_{s,c},\alpha_{s,m},\hat{V}_{s,0},\hat{\rho},t_D,a_D$ & 13 \\ \bottomrule					
			\end{tabular}
		\end{threeparttable}
	\end{center}
\end{table}

\subsection{Model fitting}
We fitted each model to the trial by trial values of the decision threshold for each participant separately. In all models, the threshold for condition $s$ at trial $k$ is modeled as a Gaussian random variable with mean $m_{s,k}(t)$ and variance $\sigma^2_{s,k}$. For a given set of values of the parameters for each model, we can compute these quantities for all $k$ and $s$. The likelihood of observing the value $a_k$ for the decision threshold for a participant at trial $k$ given that the trial was from condition $s$ is:

\begin{equation}
\label{eq:ModelFit_1}
l_k=\Pr[a_k|s,k,m_{s,k}(t),\sigma^2_{s,k},t_k]=\frac{1}{\sqrt{2\pi\sigma^2_{s,k}}}\exp\bigg(-\frac{[a_k-m_{s,k}(t_k)]^2}{2\sigma^2_{s,k}}\bigg)
\end{equation}   

Notice that we have used $m_{s,k}(t_k)$, the value of the decision threshold at time $t_k$, where $t_k$ is the participant's reaction time in trial $k$. Also, although not explicitly mentioned in this equation, we use $m_{s,k}(t_k)>0$ if the participant's response was correct and $m_{s,k}(t_k)<0$ if the response was incorrect in rial $k$. Therefore, the observed data for each participant has the form ${(y_k,t_k,a_k)}$, $k=1,2,...,N$, where $y_k$ is the participant's response (left or right). For each participant and each model we found the values of the parameters of the model such that the likelihood of the observed data was maximized. To find these values we used the differential evolution method for optimization (DEopim package in R, \cite{mullen_deoptim:_2011}).

\section{Results}
\subsection{Optimal performance}
In this section, we compute the optimal decision thresholds which result in the maximum possible average reward rate for the two experiments. Previous studies have used dynamic programming to obtain the optimal decision thresholds \cite{busemeyer_psychological_1988,frazier_sequential_2008,rao_decision_2010,drugowitsch_cost_2012}. To do this, both time and the range of possible values for the accumulated information are discretized. Then it is assume that at each time-step and value of the accumulated information, the participant has to make one of two possible actions: continue accumulating information or stop and make a decision. There is a cost for each time step that the decision is not made. This results in a discrete time and state MDP for which the optimal policy can be found using dynamic programming.  

Here, we take another approach for computing the optimal thresholds. Since in all computational models that we consider, the time-varying boundaries are modeled as a Weibull function, for each experiment we find the values of the parameters of a Weibull function that maximizes the reward rate. Specifically, for each set of values of the parameters of the Weibull function, we simulated the canoe movement path 10000 times and computed the reward and reaction time for each simulated path. Then the reward rate was computed as the average value of the reward divided by the average value of the reaction time in these simulations. We found the optimal values of the parameters using differential evolution method for global optimization (DEopim package in R, \cite{mullen_deoptim:_2011}).

\begin{figure}[!h]
 	\centering
 	\includegraphics[width=.5\linewidth]{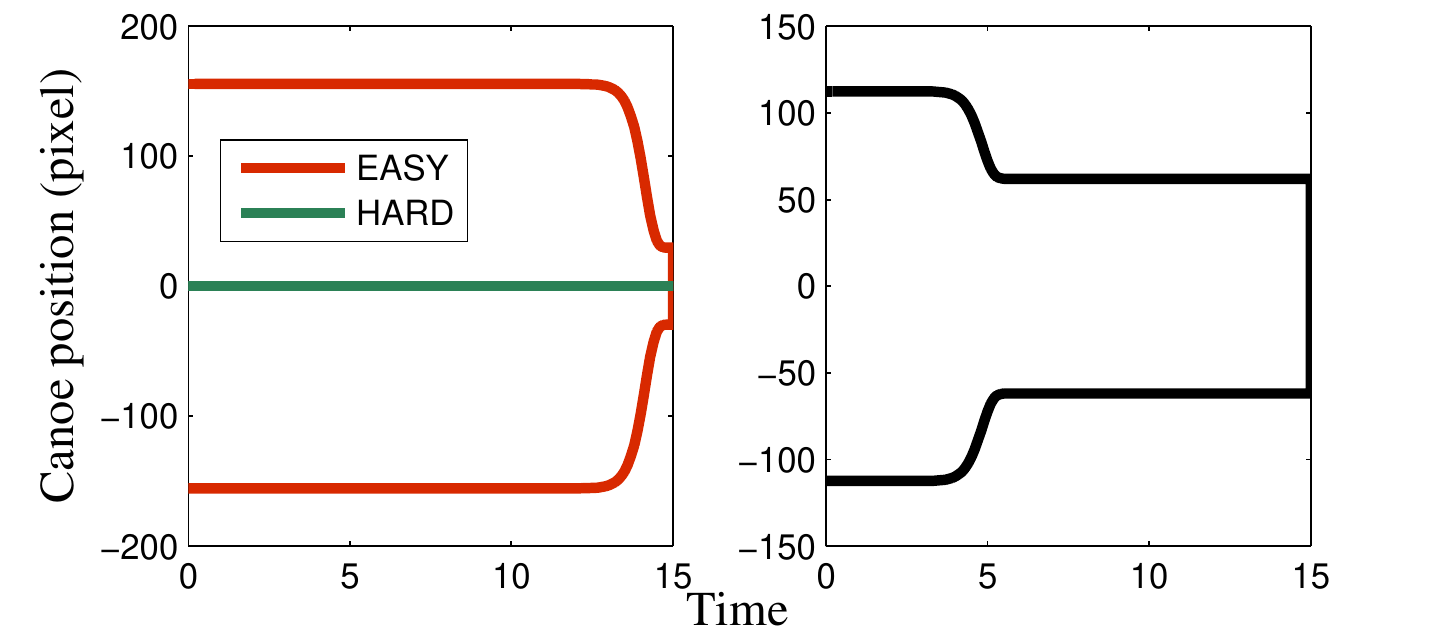}		
 	
 	\caption{{\bf Optimal decision thresholds.} Left: The optimal decision thresholds for the easy and hard conditions in Experiment A. The threshold for the hard condition is 0. Right: the single optimal threshold for both easy and hard trials in Experiment B.} \label{fig:OptimalThresholds}
 \end{figure}

The optimal thresholds for Experiments A and B are shown in Figure \ref{fig:OptimalThresholds}. The optimal threshold for the hard trials in Experiment A is zero. For the easy trials, the optimal threshold remains constant (at about 155 pixels) up to about 13 secs and collapses rapidly afterwards. This collapse is because there is a 15 secs deadline in each trial and if the participant does not respond before this time the trial is considered as incorrect. Therefore, the thresholds collapse to insure that a respond will be given in each trial. The constant optimal threshold is interesting. If there is only one type of trials, based on Wald's famous theorem on sequential probability ratio test, the optimal threshold is constant \cite{wald_optimum_1948,bogacz_physics_2006}. Our results suggest that even if there are more than one type of trials, intermixed randomly, and if it is possible to use a different threshold for each type of trials (by presenting a cue in Experiment A), then the optimal threshold for each condition is also constant\footnote{We computed the optimal thresholds for many different values of the experiment's parameters (values of the rewards, difficulty levels, delay penalties and so on) and for all of them the optimal thresholds were time-constant.}. 

In section Materials and Methods, we proposed two models for Experiment B. In the first model the participant adopts one time-varying threshold for both easy and hard trials, while in the second model the participant tries to detect the difficulty of the trial and then sets the decision thresholds accordingly. The right panel of Figure \ref{fig:OptimalThresholds} shows the single optimal threshold for the first model. As it can be seen, the optimal threshold is time-decreasing. Intuitively, using this form of the decision threshold looks reasonable: if the accumulated information has not reached the decision threshold after some time, the trial is more likely to be a hard trial and so it is not worth it to spend much more time on it, and therefore it is reasonable to decrease the decision threshold. With this threshold the average reward rate is 0.905.

In the second model, the participant accumulates information up to a time $t_D$. If the canoe position reaches a threshold $a_D$ before this deadline, the participant detects that the trial is easy and sets her decision threshold at $\pm a_E$. Otherwise, she will use $\pm a_H$ as the thresholds in that trial. We computed the values of the parameters $t_D,\,a_D,\,a_E$ and $a_H$ for which the reward rate is maximized in Experiment B. The optimal values of the parameters are: $t_D=1.5,\,a_D=135,\,a_E=135,\,a_H=90$. For these values the average reward rate is 0.88. Therefore, the maximum average reward rate obtained by the first model is larger than the second model. In other words, detecting the difficulty and then setting the decision thresholds does not have any advantage over using the same decision threshold for both easy and hard trials. However, as we will see, most of the participants used the second strategy.

\subsection{Behavioral results of experiment A}
\begin{figure}[!h]
	\centering
	\includegraphics[width=1\linewidth]{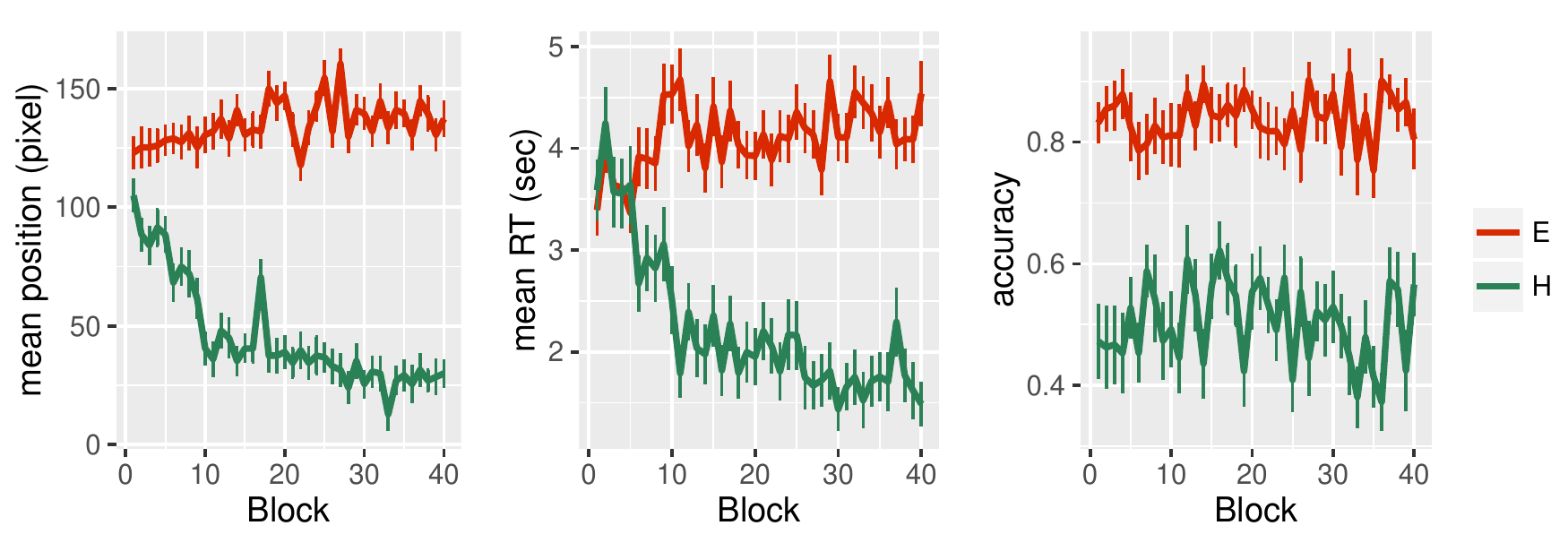}		
	
	\caption{{\bf Results of Experiment A.} Left: mean position of the canoe at the time the participant made her decision. Middle: mean RT. Right: accuracy. E: easy trials. H: hard trials. Each curve is created by computing the average of the corresponding quantity over 26 participants. The bars indicate standard error.} \label{fig:Results_A}
\end{figure}

This experiment consisted of 40 blocks of trials. We recorded participants' choice, reaction time and the canoe position at the time the participant responded (the decision threshold) in each trial. These quantities as functions of the block number, averaged across all 26 participants, are shown in Figure \ref{fig:Results_A}. As it can be seen in the left panel of this figure, on average the participants learn to decrease their threshold for responding in the hard trials. Also, it seems that the threshold in the easy trials increases slightly. Mixed-effect regression analysis with block number and condition as regressors and considering block number as the random effect, showed that the change in the decision threshold (as a function of the block number) in the easy condition is not significant ($p=0.3508$) while it is significant in the hard condition ($p<0.0001$). Also, the difference between the two conditions is significant ($p<0.0001$). 

\begin{figure}[!h]
	\centering
	\includegraphics[width=.5\linewidth]{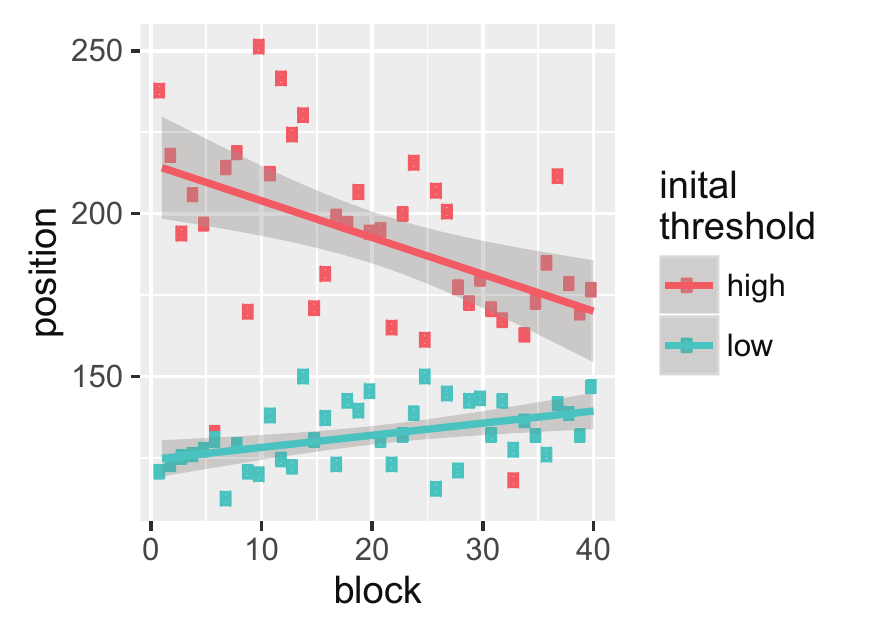}		
	
	\caption{{\bf Decision threshold saparated by initial value.} The decision threshold for the easy trials in Experiment A as a function of the block number is shown separated by the average value of the decision threshold in the first block. High: for these participants the average decision threshold in the first block was higher than 155 pixels. Low: for these participants this value was lower than 155 pixels. The lines show the best fitted linear model to the data.} \label{fig:Results_A_SeperatedByInitiThd}
\end{figure}

In the previous section, we showed that the optimal value of the decision threshold is 155 and 0 pixels for the easy and hard trials, respectively. As it can be seen in Figure \ref{fig:Results_A}, for the hard trials the participants' thresholds were much higher than the optimal value at the beginning of the experiment and decreased by learning. For the easy trials, for 5 of the participants the average of the decision threshold in the first block was higher than 155 pixels, and for 21 participants this value was lower than 155 pixels. Figure \ref{fig:Results_A_SeperatedByInitiThd} shows the average value of the decision threshold in the easy trials in each block for these two groups of participants. The best fitted linear model to this data is also shown in this figure. The slope of the line is -1.12 ($t(38)=3.31,\, p=0.002$) for the group with higher than optimal initial threshold, and 0.12 ($t(38)=3.06,\, p=0.004$) for the group with lower than the optimal initial threshold. Therefore, the direction of the change in the decision threshold is toward the optimal value for both groups of the participants. 

\begin{figure}[!h]
	\centering
	\includegraphics[width=.8\linewidth]{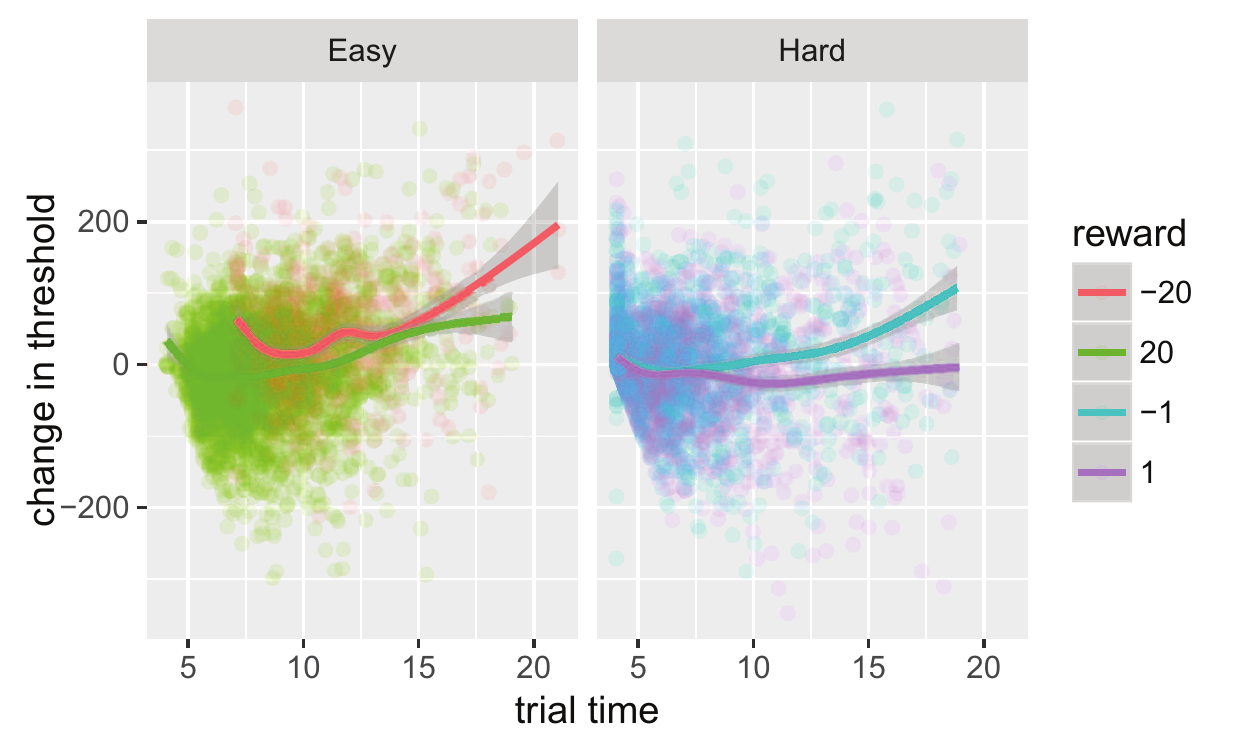}		
	
	\caption{{\bf Change in decision threshold after each trial.} If we denote the value of the decision threshold at trial $k$ by $a_k$, each point in this figure represents $a_{k+1}-a_{k}$. A positive value of change means that the participant has increased her decision threshold. Trial time is the total time of a trial. Here, $k$ is not the actual trial index. Instead, it indexes the trials from the same condition. The data points are plotted separately for positive and negative values of reward in each condition. The rewards in the easy conditions are $\pm 20$ and they are $\pm 1$ for the hard trials. See the text for more details.} \label{fig:A_change_in_threshold}
\end{figure}

Our main focus in this paper is on how the participants adjust their decision threshold after each trial. Specifically, how the decision threshold is adjusted as a function of the rewards and the time of a trial. Figure \ref{fig:A_change_in_threshold} shows the changes from trial to trial in the decision threshold as a function of the achieved reward and the total trial time (including the delay penalty, ITI and so on). Let $a_k$ denote the value of the decision threshold at trial $k$. Each point in this figure represents $a_{k+1}-a_{k}$. Here, $k$ is not the actual trial index. Instead, it indexes the trials from the same condition. For example if a participant experienced an easy trial for the first time at trial 10 and for the second time at trial 15, then $a_{2}-a_{1}$ for the easy trials will be the difference between the threshold values in these two trials. Any conclusion from this figure should be drawn with cautious because any observed effect could be the results of several factors. We will attempt to disentangle these effects using computational modeling in the next section. However, it is worth mentioning at least one effect observed in this figure. For both hard and easy trials, the change in the decision threshold after receiving a negative reward is higher than the positive reward. In other words, the participants increase their decision thresholds after committing an error. This is consistent with the well-known post-error slowing effect \cite{laming_choice_1979,smith_slowness_1995}.  

Figures 1 and 2 in S1 Text show the decision threshold for each participant during the experiment. As it can be seen in these figures, there is a lot of individual differences. One question that arises is how many participants have eventually learned the optimal decision thresholds. To answer this question, for each participant we tested if the decision thresholds in the last five blocks were significantly different from the optimal values. For the hard trials, we used a binomial test on the number of trials in which the decision threshold was less than 45 pixels\footnote{In each step of the Markov chain governing the canoe movement, the canoe moves 45 pixels in one direction. This is why we have used 45 pixels to test if the participant has used 0 decision threshold for the hard trials.}. With the significant level of 0.05, this test was significant for 14 out of 26 participants. Therefore, 12 participants did not learn to reduce their decision threshold to 0 in the hard trials by the end of the experiment. For the easy trials, we performed a t-test with the null hypothesis that the decision threshold is not different from 155. The test was not significant for 11 participants which means that the decision threshold of these participants were not significantly different from the optimal value. Also, 8 participants used lower than optimal and 7 participants used higher than optimal threshold in the last 5 blocks of trials. 

\subsection{Comparison of computational models of experiment A}
We attempted to characterize the participants' learning by fitting 10 computational models to the behavioral data. The models are summarized in Table \ref{tab:models_A}. All models were fitted to each participant's data separately. We do not expect that all participants use the same learning mechanism. Some participants may not adjust their decision threshold during the experiment. Therefore, it is important to examine which model can explain each participant's data the best.

\begin{figure}[!h]
	\centering
	\includegraphics[width=.5\linewidth]{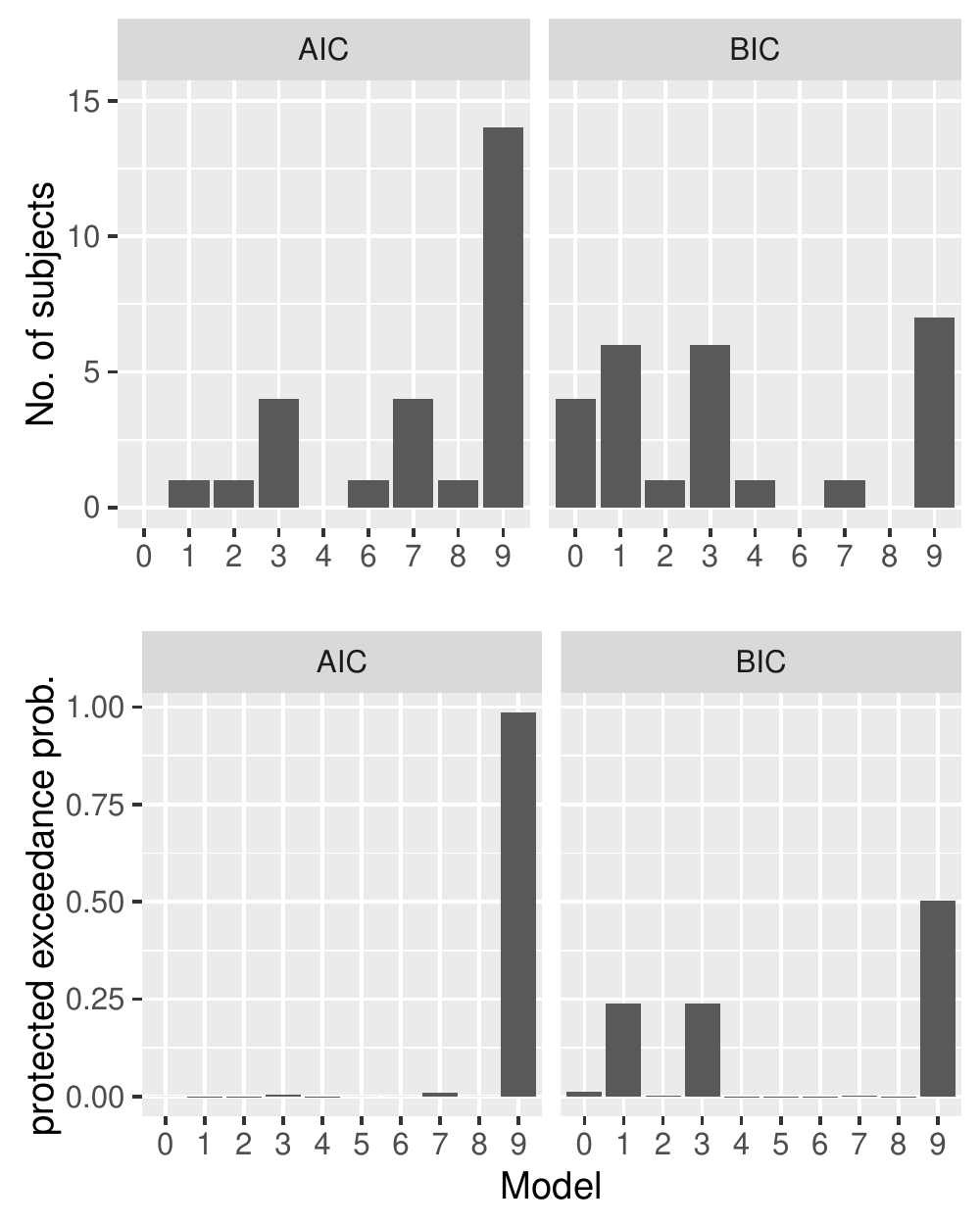}	
	\caption{{\bf Comparison of the computational models of experiment A.} Top row: the number of participants (out of 26) for which each model was the best among all 10 models. Bottom row: the protected exceedance probabilities for each model. The comparisons are based on AIC  and BIC in the left and right panels respectively.} \label{fig:ModelComparison_A}
\end{figure}

\begin{figure}[!h]
	\centering
	\includegraphics[width=.8\linewidth]{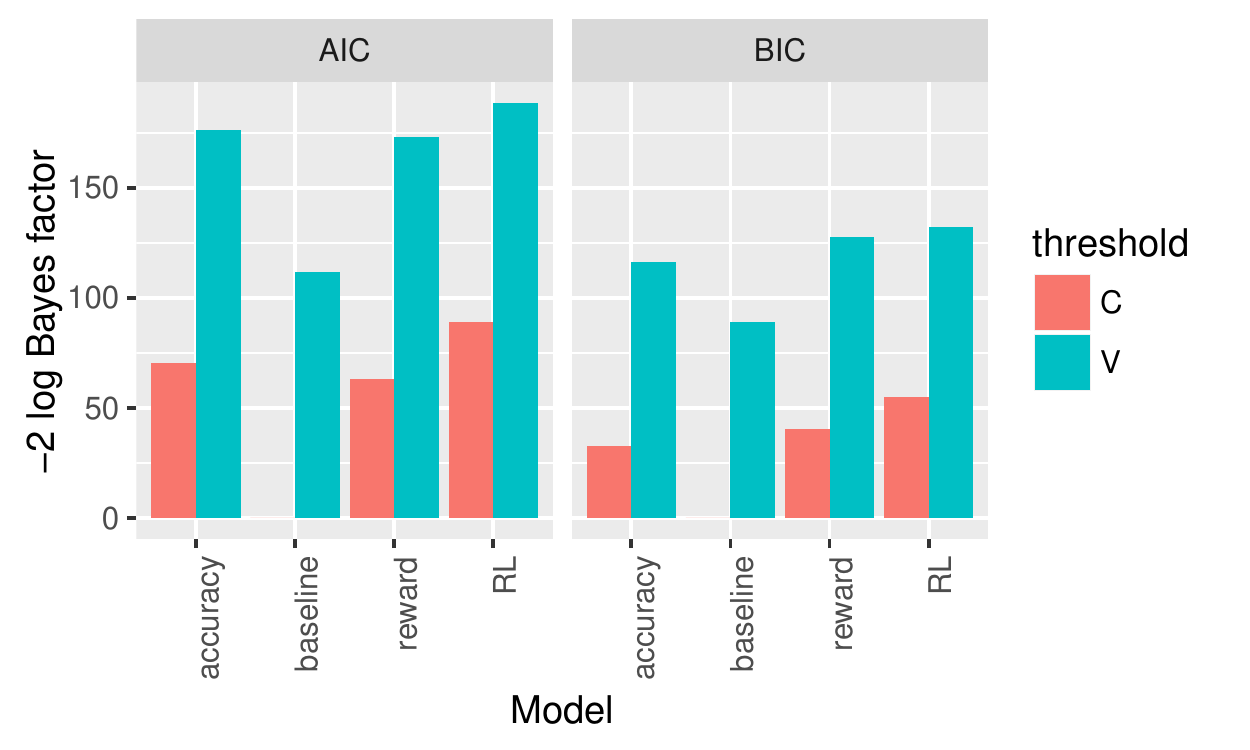}	
	\caption{{\bf Bayes factors for each model versus Model 0 in Experiment A.} The vertical axis is $-2\log\,Bayes\,Factor$ estimated by the difference between the AIC (left) and BIC (right) of each model and Model 0. The color of the bars indicates if the model uses time-constant (C) or time-varying (V) decision thresholds. The models are grouped into 4 categories. Baseline: models 0 and 1. Accuracy: models 2 and 3. Reward: models 5 and 6. RL: models $RL_C$ and $RL_V$. See Table \ref{tab:models_A}. As it can be seen, for each category, the time-varying version performed much better than the time-constant version. Also, all models performed significantly better than Model 0.} \label{fig:BayesFactor_A}
\end{figure}

The results of the model fitting and comparison are depicted in Figures \ref{fig:ModelComparison_A} and \ref{fig:BayesFactor_A} and Table \ref{tab:gof_A}.The top row of Figure \ref{fig:ModelComparison_A} shows the number of participants for which each model has been the best model based on both the Bayesian information criterion (BIC) and the Akaike information criterion (AIC). Based on both measures of the goodness of fit, the RL model with time-varying decision thresholds is the best model for most of the participants.

One way to compare the models is to treat BIC/AIC as the log model evidence for each participant and investigate if there is a significant difference between the BIC (AIC) for a pair of models. The average (across participants) difference between BIC of Model 0 and Model 9 (RL model with time-varying thresholds) was 131 which was significantly greater than 0 ($t_{25}=4.23,p=0.0002$). This difference was $BIC_1-BIC_9=42.96$  ($t_{25}=2.56,p=0.017$) for Model 1 versus Model 9, and $BIC_3-BIC_9=4.18$  ($t_{25}=0.77,p=0.44$). However, $AIC_3-AIC_9=15.50$ ($t_{25}=2.84,p=0.008$). In words, based on AIC Model 9 performed significantly better than Model 3, but the difference between the models is not significant based on BIC. The average of the difference between the BIC/AIC of each model to Model 0 is shown in Figure \ref{fig:BayesFactor_A}. It can be seen that all models perform strongly better than this baseline model.

We can compute this difference for each pair of the models. However, this needs many pairwise comparisons. Another problem is that the results based on AIC and BIC are not consistent to some extent: based on AIC, there is much stronger evidence that model 9 is the best model. Part of the problem is that we are comparing 10 models and this makes the comparisons to be affected more by possible outlier values in BIC or AIC. A more sophisticated approach for comparing this number of models have been proposed by \cite{stephan_bayesian_2009}. In this method, the models are considered as random variables. The probabilities that the data of a participant chosen at random is generated by each model form a multinomial distribution. The parameters of this distribution are described by a Dirichlet distribution. Stephan et al. proposed a simple variational algorithm to estimate the parameters of this distribution (see \cite{stephan_bayesian_2009} equation 14). Given the parameters of the Dirichlet distribution, it is possible to compute the probability that a model is more likely than any other model. This is called the \textit{exceedance probability (EP)}. To compute these probabilities, it is necessary to have an estimate of the marginal likelihood of each model $m$ for the data set $D$, i.e. $p(D|m)$. The EPs for all 10 models, using BIC and AIC as the estimate for $\log(p(D|m))$, are shown in the bottom row of Figure \ref{fig:ModelComparison_A}\footnote{To compute PEs, we used the MATLAB code developed by Samuel J. Gershman and available at . See \cite{gershman_empirical_2016} for more details.}. Based on both AIC and BIC as the estimate, the RL model with time-varying thresholds is the most likely model. The EP values based on AIC, strongly support this model. The EP value based on BIC is 0.53 for this model, and 0.21 and 0.23 for models 1 and 3, respectively.  

\begin{table}
		\centering
		\caption{Goodness of fit of computational models of Experiment A}
		\label{tab:gof_A}
	\begin{center}
		\begin{threeparttable}					
			\begin{tabular}{lccc}
				\toprule
				\textbf{Model} & \textbf{NLL} & \textbf{AIC} & \textbf{BIC} \\ \midrule
				 0 & 1735(51.7) & 3478 & 3493 \\
				 1 & 1673(47.3) & 3367 & 3404 \\
				 2 & 1697(49.5) & 3415 & 3453 \\
				 3 & 1636(46.2) & 3305 & 3366 \\
				 4 & 1632(45.8) & 3301 & 3369 \\
				 5 & 1690(48.7) & 3408 & 3461 \\
				 6 & 1631(45.8) & 3302 & 3377 \\
 				 7 & 1632(45.4) & 3300 & 3368 \\
				 $\mathrm{RL_C}$ & 1681(48.8) & 3389 & 3438 \\
				 $\mathrm{RL_V}$ & 1626(45.7) & 3290 & 3361 \\ \bottomrule
			\end{tabular}
			\begin{tablenotes}
				\footnotesize \textit{note:} The numbers in the parentheses are standard errors. NLL=negative log likelihood. 
			\end{tablenotes}
		\end{threeparttable}
	\end{center}
\end{table} 

An important point that is consistent among all these analyses is that for each class of models, the time-varying version of the model performs better than the time-constant version. The PEs for all models with time-constant thresholds, based on both AIC and BIC, is almost zero. If a participant is using time-varying thresholds but we fit a time-constant threshold model to her data, the residuals between  the observed canoe position (at the time that the participants has responded) in each trial and the predictions of model will be a function of the reaction time in the trials. For example suppose that the participant has used time-decreasing decision thresholds. In this case, the observed canoe positions for trials with short reaction times will be larger than those with longer reaction times. Now if we fit a model with time-constant threshold, the best fitted value of the decision threshold will be smaller than the true decision threshold for short reaction times, and larger than the true decision threshold for longer reaction times. Therefore, the residuals between the observed canoe positions in each trial and this best fitted value will be a function of the reaction time.  

\begin{figure}[!h]
	\centering
	\includegraphics[width=.5\linewidth]{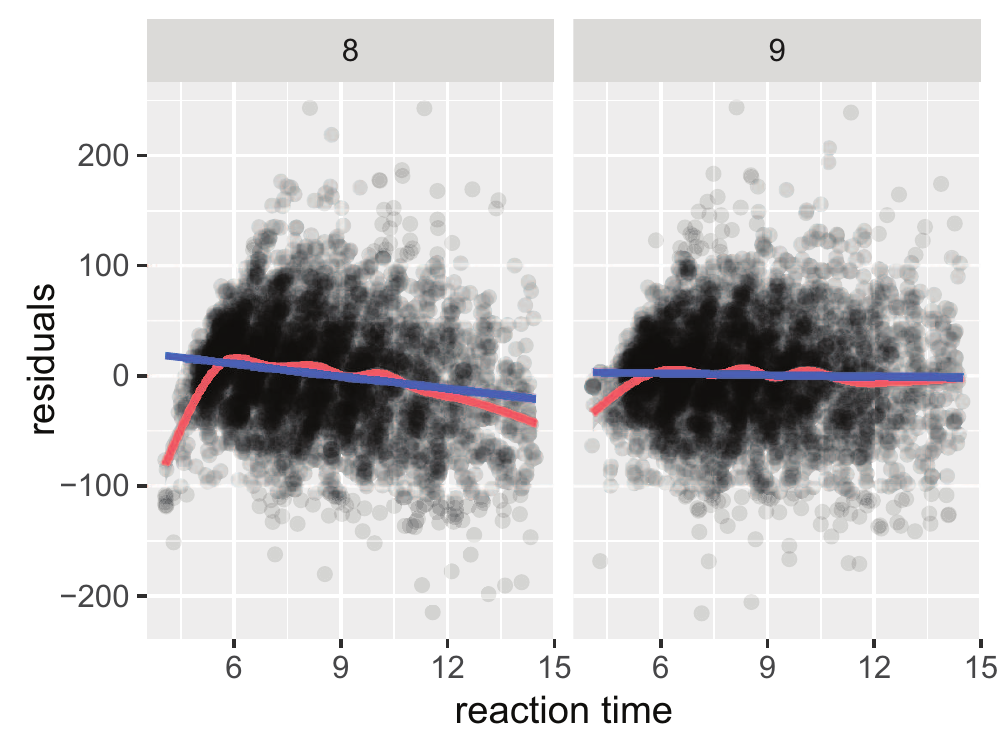}	
	\caption{{\bf Residuals between observed value of decision threshold and predicted values by Models 8 and 9.} The data is for all participants and all trials. The blue line shows the best fitted linear model to the data and the red line shows the fitted spline smoother. The residuals for Model 8 decrease as reaction time increases.} \label{fig:Residuals_model8vs9}
\end{figure}

 The residuals between the observed canoe position in each trial and the predictions of models 8 and 9 as a function of the reaction time are shown in Figure \ref{fig:Residuals_model8vs9} (for all trials from all participants). A fitted linear model and spline smoother are also shown in this figure. Model 8 and 9 are the time-constant and time-varying variants of the RL model. Among models with time-constant threshold, Model 8 has the lowest AIC and BIC (see Table \ref{tab:gof_A}). As it can be seen, the residuals for model 8 are a function of the reaction time while this is not the case for model 9. The slope in the fitted linear model for the residuals of Model 8 is -3.73 and significantly lower than zero ($p<2e-16$). For the residuals of Model 9 the slope is -0.468 and is not significantly different from zero ($p=0.15$). This provides another evidence favoring the models with time-varying thresholds over the models with time-constant threshold.
 \newline

One reason that the results from AIC and BIC are partly inconsistent could be that the number of the free parameters of the models is large relative to the number of the trials (between 350-450 trials for different participants). To examine this possibility more, we ran 2 more male participants for three sessions of Experiment A. The first session consisted of 40 blocks and the other two sessions consisted of 35 blocks. Each session was held in one day and there were not more than 2 days gap between the sessions. All other task parameters were exactly the same as for the one-session version of the experiment.

We fitted models 1, 3 and 9 to the data of these participants. participants 1 and 2 experienced 957 and 1026 trials during the three sessions, respectively. The values of the negative log-likelihood, AIC and BIC are presented in Table \ref{tab:gof_A_2}. Interestingly, for both participants the RL model with time-varying thresholds is the best model based on both AIC and BIC.

\begin{table}
	\centering
	\caption{Goodness of fit of models 1,3 and 9 for three session version of Experiment A}
	\label{tab:gof_A_2}
	\begin{center}
		\begin{threeparttable}					
			\begin{tabular}{llccc}
				\toprule
				\textbf{participant} & \textbf{Model} & \textbf{NLL} & \textbf{AIC} & \textbf{BIC} \\ \midrule
				   & 1 & 4887 & 9794 & 9843 \\
				1 & 3 & 4815 & 9666 & 9754 \\
				   & $\mathrm{RL_V}$ & 4765 & 9569 & 9662 \\ \midrule
				   & 1 & 4765 & 9550 & 9599 \\
			    2 & 3 & 4434 & 8905 & 8994 \\
				   & $\mathrm{RL_V}$ & 4402 & 8843 & 8936 \\ \bottomrule
			\end{tabular}

		\end{threeparttable}
	\end{center}
\end{table} 

\begin{figure}[!h]
	\centering
	\includegraphics[width=.5\linewidth]{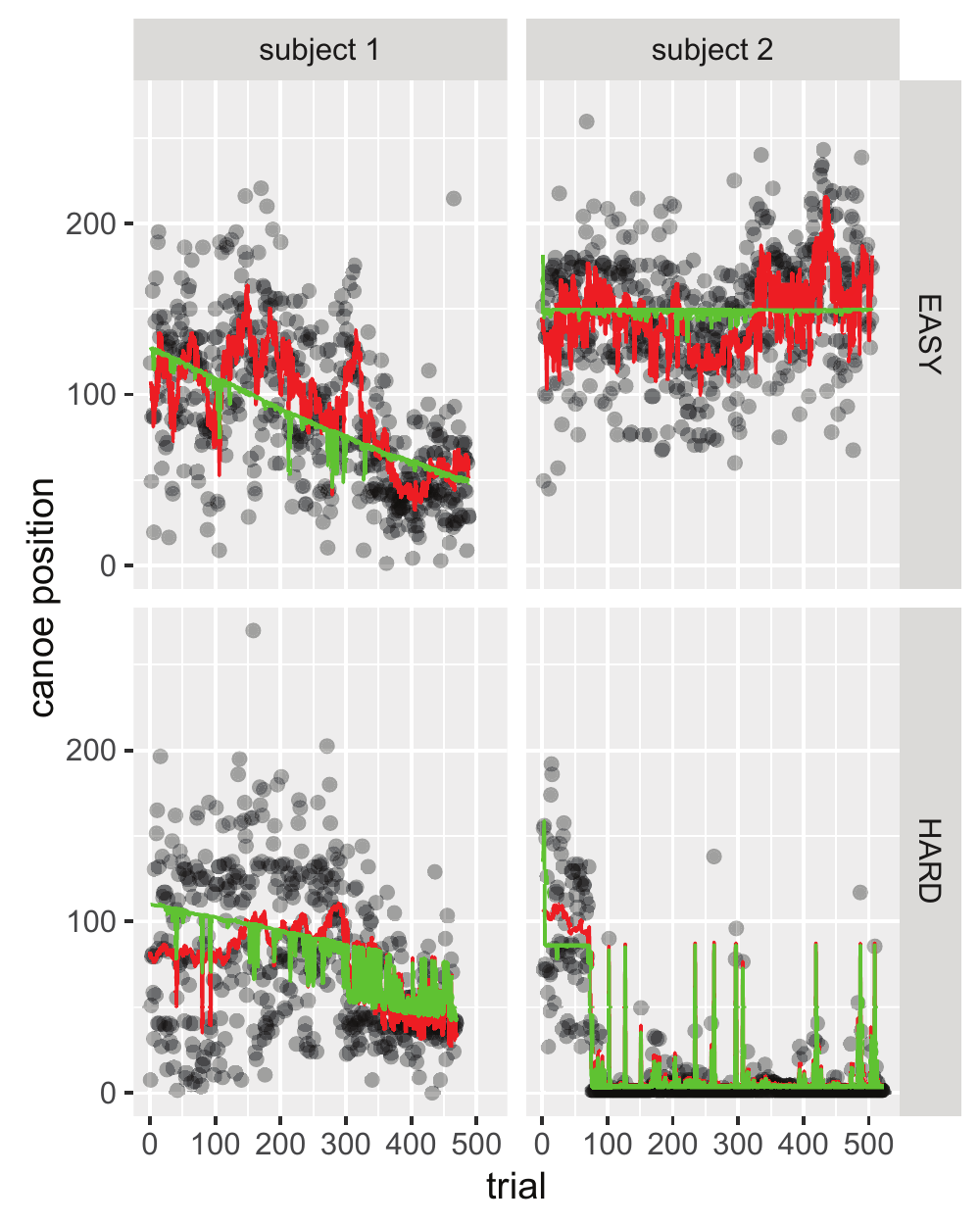}	
	\caption{{\bf Observed and predicted values of decision threshold in 3-sessions of Experiment A.} The two participants participated in three sessions of Experiment A. The black circles are the observed values of the decision threshold. The red and green curves are the predictions of Model 9 and Model 3 of the decision thresholds, respectively.} \label{fig:Models_Predict_3session_A}
\end{figure}

The observed values of the canoe position in each trial (observed decision threshold), together with the predicted values of the decision threshold from Models 3 and 9 are shown in Figure \ref{fig:Models_Predict_3session_A} for the two participants in this experiment. As it can be seen, Model 9 fits the data much better. Specifically, for the easy trials, Model 3 tries to capture the ``general trend" in the data and has not been able to capture more subtle changes in the decision thresholds. We observed similar patterns in the behavior of the fitted thresholds in Model 3, for some of the participants in the one-session version of the experiment (For example compare the left most column of Figures 3 and Figure 5 in S1 Text). By investigating Figures \ref{fig:Models_Predict_3session_A} and Figures 3 and 4 in S1 Text, it seems that this model, in general, does not have enough complexity to be able to capture many of the observed patterns in the data.   

\subsection{Behavioral results of experiment B}
This experiment consisted of 35 blocks of trials. The crucial difference between this experiment and Experiment A is that here there is no cue indicating the condition of the upcoming trial. Therefor, if the participant sets her decision threshold at the beginning of each trial, she must use the same decision threshold for both easy and hard trials. On the other hand, the participant may use a mechanism to detect the condition of the trial first and then sets her decision threshold.

\begin{figure}[!h]
	\centering
	\includegraphics[width=1\linewidth]{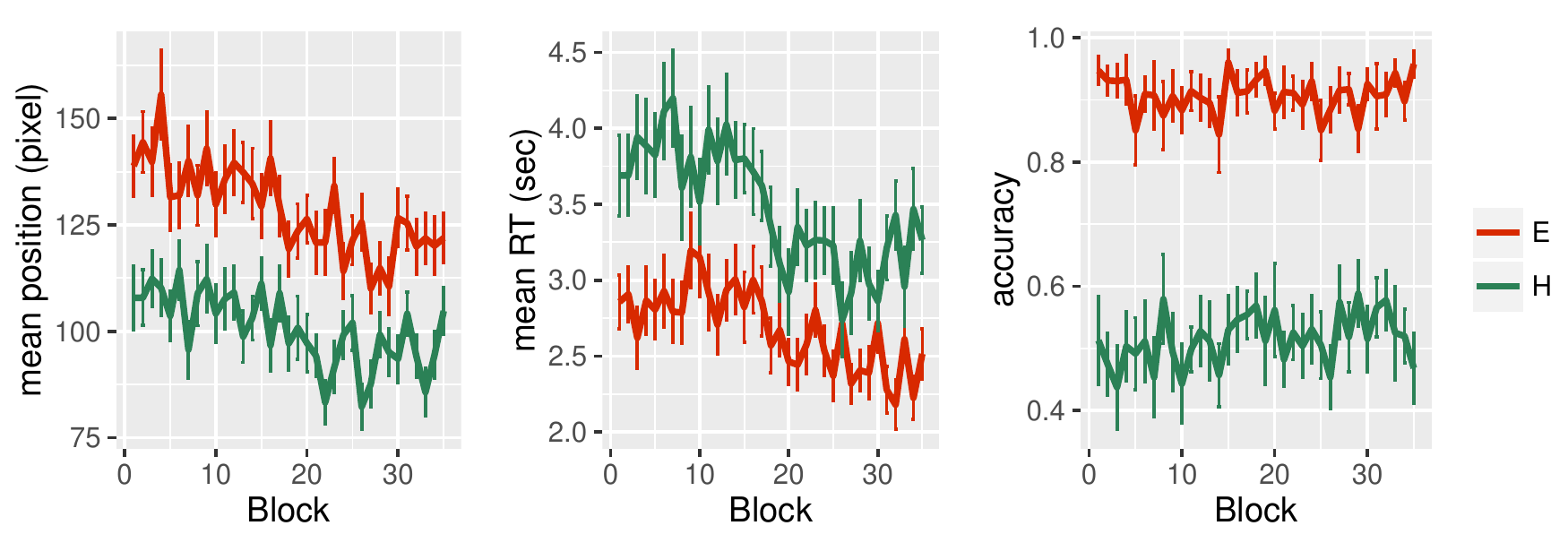}	
	\caption{{\bf Results of Experiment B.} Left: mean position of the canoe at the time the participant made her decision. Middle: mean RT. Right: accuracy. E: easy trials. H: hard trials. The bars indicate standard error.} 
	\label{fig:Results_B}
\end{figure}

The decision thresholds, reaction times and accuracy, averaged across all 20 participants, are shown in Figure \ref{fig:Results_B}. As it can be seen in this figure, the decision threshold decreases in both easy and hard trials. Interestingly, on average, the decision thresholds for the easy and hard conditions are different. The linear mixed-effect analysis of the decision threshold (with block number and condition as regressors and considering block number as the random effect) showed that the decision thresholds were significantly different in the easy and hard conditions ($p<0.0001$). Also, the reduction in the decision threshold is significant in both conditions ($p=0.0091$), but it is not different between the conditions ($p=0.0829$).

\begin{figure}[!h]
	\centering
	\includegraphics[width=.5\linewidth]{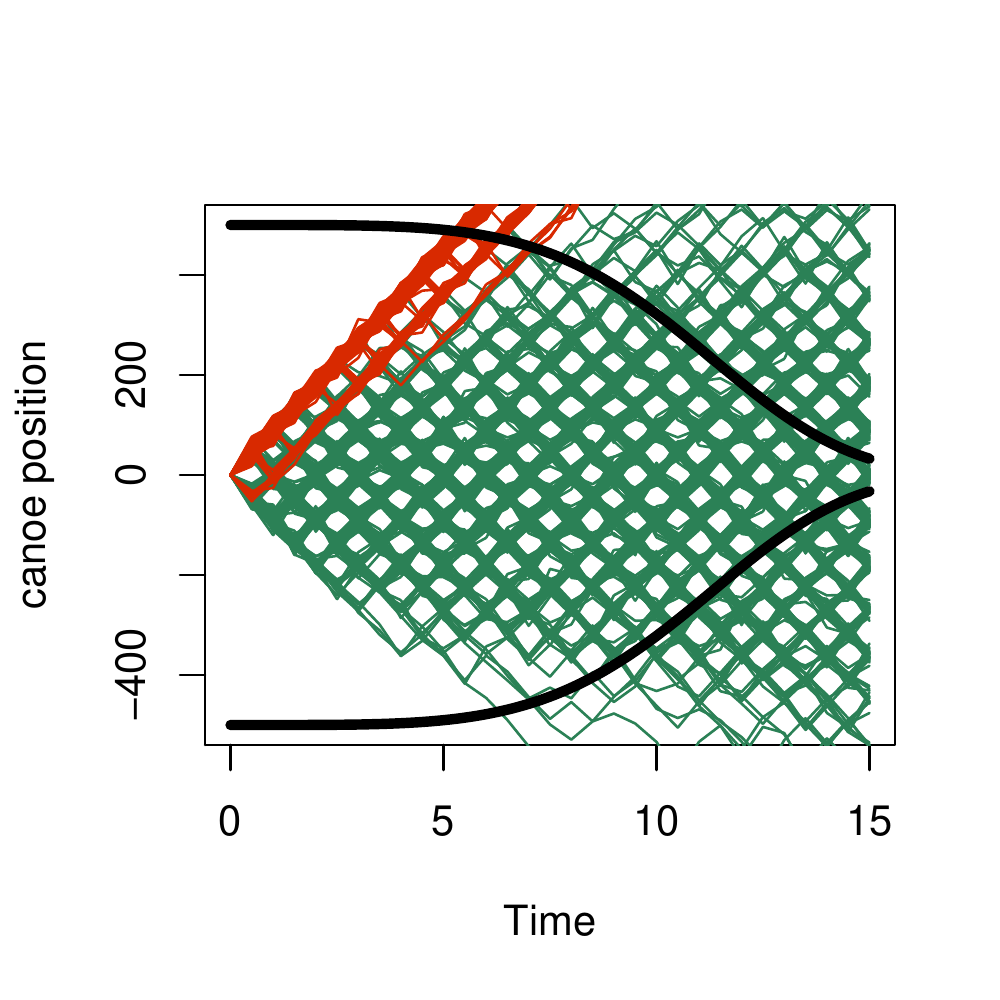}	
	\caption{{\bf Simulated canoe paths.} The red paths correspond to the easy trials with $P_0=0.99$ and the green paths correspond to the hard trials with $P_0=0.5$. To make the figure visually clearer, we have added Gaussian noise to each path. The black curves are obtained from a Weibull function (Equation \ref{eq:RL_models_06}) with parameters $\psi=500, \psi \prime=-200,\lambda=12,\phi=4.5$. Most of the red paths reached the decision threshold around 5 secs when the decision threshold is large, while most of the green paths reached the decision threshold later when its value has decreased. This shows that even with a single time-decreasing decision threshold, the observed values of the decision threshold for the hard trials will be lower than in the easy trials.} 
	\label{fig:Simulate_B_WeibullThd}
\end{figure}

It is important to note that the difference between the decision thresholds for the easy and hard trials, observed in Figure \ref{fig:Results_B}, does not necessarily imply that the participants are using two different decision thresholds for these two conditions. A single time-decreasing decision threshold for all trials can also explain the pattern observed in this figure. Since in the hard trials the probability that the canoe moves to the correct direction is lower, on average it takes longer for the canoe to get away from the center of the screen. Therefore, if a participant uses a time-decreasing threshold, in most of the hard trials, the canoe will not reach the threshold position in a short course of time, and so the mean RT will be higher and the decision threshold will be lower for the hard trials. To clarify this, the result of simulating the canoe position using two values of $P_0$ (the probability that the canoe moves to the correct direction in each time step), together with an example of a time-decreasing threshold is shown in Figure \ref{fig:Simulate_B_WeibullThd}. For the red paths, $P_0=0.99$ and for the green paths $P_0=0.5$. As it can be seen, most of the red (which correspond to the easy trials) paths reach the threshold around 5 secs, when the value of the decision threshold is high. Most of the green paths, on the other hand, reach the threshold later when the decision threshold has decreased. This shows that the patter of RT and decision thresholds observed in Figure \ref{fig:Results_B} can be accounted for by a single time-decreasing decision threshold. Therefore, we need more sophisticated methods to distinguish between a single-threshold and two-threshold hypotheses. In the next section, we use computational modeling to investigate this question more rigorously.  

\subsection{Comparison of computational models of experiment B}
Our goal here is to compare two hypotheses. Based on the first hypothesis, the participants set their decision threshold at the beginning of each trial and before the trial starts. Since there is no cue presented at this time, the participants have to use the same decision threshold for both the easy and hard trials. This assumption has been made in previous research \cite{ratcliff_connectionist_1999,ratcliff_diffusion_2002,ratcliff_comparison_2004}. Importantly, in Experiment B, the optimal behavior is also to use one decision threshold for both types of trials. For the task parameters used for this experiment (rewards, delay penalties, inter-stimulus intervals and so on) this optimal threshold is a time-decreasing function of the elapsed time in a trial. Based on the second hypothesis, in each trial the participant first attempts to recognize the difficulty level of the current trial, and then sets different decision thresholds based on the detected difficulty level.    

To compare these two hypotheses, we fitted computational models corresponding to each hypothesis and examine which model can account for the data better. The computational models corresponding to each hypothesis are summarized in Table \ref{tab:models_B}. Models $M_{H_1}^1$, $M_{H_1}^2$ and $M_{H_1}^2$ all assume that the participants use only one threshold for both easy and hard trials. The threshold is modeled as the Weibull function. The difference between these models is in the way the threshold is updated. In $M_{H_1}^1$ there is no learning mechanism, while in $M_{H_1}^2$ the experiment is modeled as an one-state SMDP and the parameters of the decision thresholds are updated using a variant of the \textit{REINFORCE} algorithm (Equation \ref{eq:RL_models_03}). In $M_{H_1}^2$ the experiment is modeled as a two-state SMDP but the available action in both states (the decision threshold) is the same. In $M_{H_2}$, the model based on the second hypothesis, the experiment is modeled as an SMDP with two states corresponding to the two types of trials, easy and hard. In each trial, if the canoe position reaches a threshold $a_D$ before an internal deadline $t_D$, the participant recognizes the trial as easy and sets her decision thresholds at $\pm a_E$. Otherwise, the trial is considered to be hard and the participant uses thresholds $\pm a_H$. The value of these thresholds are updated using an RL algorithm. See Materials and Methods section for more details.  

\begin{figure}[!h]
	\centering
	\includegraphics[width=.5\linewidth]{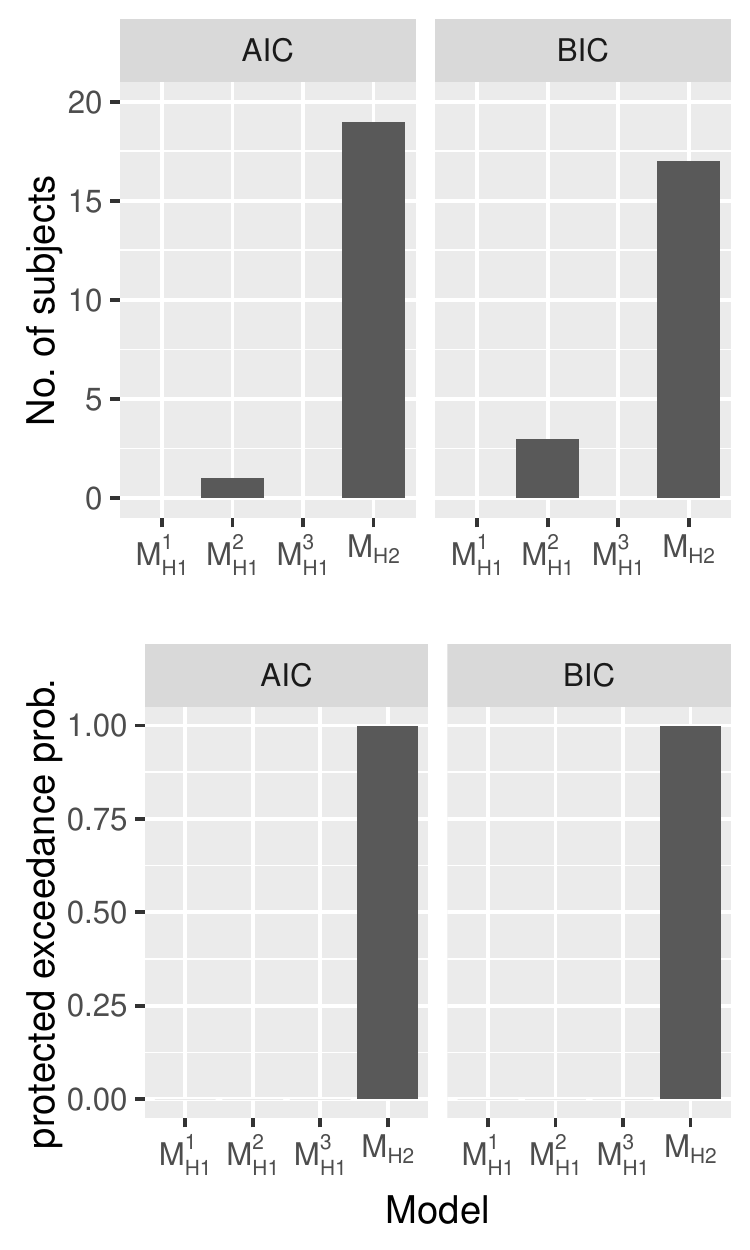}	
	\caption{{\bf Comparison of the computational models of experiment B.} Top row: the number of participants (out of 20) for which each model was the best among all 4 models. Bottom: PEs for all models. In each row, the left panel is based on AIC and the right panel is based on BIC.} \label{fig:ModelComparison_B}
\end{figure}

The number of participants for which each model was the best and the exceedance probabilities (EP) based on both AIC and BIC for the three models are shown in Figure \ref{fig:ModelComparison_B}. Model $M_{H_2}$ is the best model for 18 and 17 participants out of 20, based on AIC and BIC, respectively. The EPs based on both AIC and BIC are almost 1 for this model and 0 for the models based on the first hypothesis. NLL, AIC and BIC of the models averaged across all participants are presented in Table \ref{tab:gof_B}. These results together strongly suggest that the participants use a mechanism to detect the difficulty of the trial first and then set the corresponding decision threshold. Next, we examine the properties of the fitted model $M_{H_2}$ more.

\begin{table}
	\centering
	\caption{Goodness of fit of computational models of Experiment B}
	\label{tab:gof_B}
	\begin{center}
		\begin{threeparttable}					
			\begin{tabular}{lccc}
				\toprule
				\textbf{Model} & \textbf{NLL} & \textbf{AIC} & \textbf{BIC} \\ \midrule

				$M_{H_1}^1$ & 1862(53.8) & 3734 & 3754 \\
				$M_{H_1}^2$ & 1821(51.5) & 3661 & 3696 \\
				$M_{H_1}^3$ & 1827(52.9) & 3679 & 3726 \\
				$M_{H_2}$ & 1772(52.8) & 3570 & 3620 \\ \bottomrule
			\end{tabular}
			\begin{tablenotes}
				\footnotesize \textit{note:} The numbers in the parentheses are standard errors. 
			\end{tablenotes}
		\end{threeparttable}
	\end{center}
\end{table} 

\begin{figure}[!h]
	\centering
	\includegraphics[width=.5\linewidth]{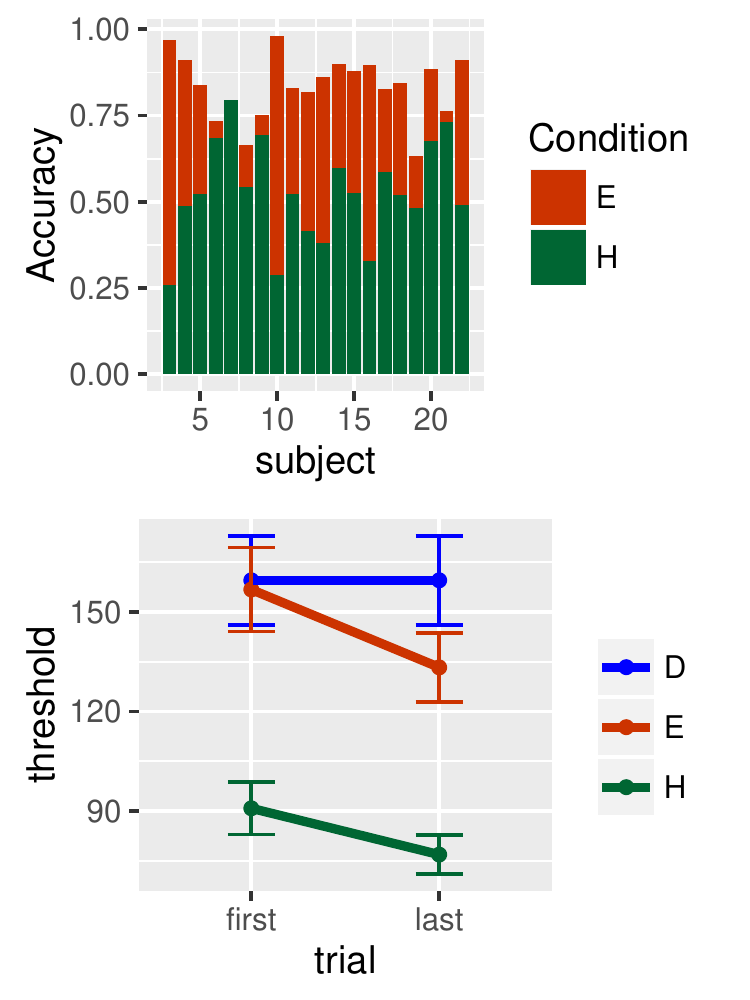}	
	\caption{{\bf Analysis of fitted model $M_{H_2}$.} Top: each participant's accuracy in detecting the difficulty level of the hard (H) and easy (E) trials. Bottom: Median of the difficulty detection threshold (D) and the thresholds in the easy (E) and hard (H) trials averaged across all participants. These quantities are computed for the first and last 20 trials for each participant to show the effect of learning.} \label{fig:Fitted_M_H2_B}
\end{figure}

The first question that may arise is how accurate participants are in detecting the trials difficulty. The top panel of Figure \ref{fig:Fitted_M_H2_B} shows the probability of correctly detecting the difficulty of a trial for each difficulty level for all participants. To generate this figure, we fitted model $M_{H_2}$ to obtain $t_D$ and $a_D$ for each participant. Then, for each participant, we computed the probabilities of correctly detecting the difficulty given the canoe path that the participant experienced in each trial and the estimated values of $t_D$ and $a_D$ for that participant. As we can see, most of the participants are much more accurate in detecting the easy trials than the hard trials. Specifically, for many participants the probability of correctly detecting a hard trial is less than 0.5. This means that these participants tended to detect a trial as easy more than hard.

The bottom panel of Figure \ref{fig:Fitted_M_H2_B} shows the average of the difficulty detection threshold ($a_D$) together with the average of the decision thresholds for the two types of trials ($a_E$ and $a_H$) at the beginning and end of the experiment. This figure is plotted as follows: For each participant, we first simulated model $M_{H_2}$ with the best fitted values of the parameters for that participant. This results in a predicted value of the mean of the decision thresholds, $m_{s,k}$, for each trial $k$. Then, we computed the median of these predicted values for the first and last 20 trials in the experiment for each participant. The figure shows the mean and the standard error of these median values across all participants. It is important to note that in $M_{H_2}$ we assume that the participants do not adjust the difficulty detection threshold and so its value is the same for the first and last 20 trials in the figure. As it can be seen in this figure, the mean of the Gaussian representing the threshold in the easy trials, $m_{E,k}$, is much larger than that of the hard trials, $m_{H,k}$. The mean in both conditions decrease with experience. Also, at the beginning of the experiment, the value of $m_{E,k}$ is close to $a_D$ and it becomes smaller than $a_D$ by experience.

The estimated value of the parameter $t_D$ (the internal deadline for detecting trial as easy) is 5.37 (with standard error of 0.55). This value is larger than the participants' average reaction time (middle panel of Figure \ref{fig:Results_B}). Together, these results show that the participants used a conservative strategy. By using a large values of $t_D$ and $a_D$ they detected most of the trials as easy trial and used higher decision thresholds for those trials. Given that the stakes for the easy trials were higher than the hard trials, this strategy is reasonable.

\section{Discussion}
When a sequence of decisions should be made during a limited time interval, the total outcome depends not only on the outcome of each decision, but also on the time spent on each decision on average. If the decisions have different properties, the decision maker should decide how much time to spend on each decision in order to achieve the maximum outcome. Little is know about how human and animals allocate limited time to decisions with different properties.

In this paper, we reported the results of two experiments to investigate this question. In both experiments, the total duration of the experiment was fixed and the number of the trials that a participant could experience depended on her speed in responding. Two types of trials were intermixed randomly in each block of the experiments: easy trials with larger absolute value of positive and negative rewards (for correct and incorrect decisions), and hard trials with smaller values of positive and negative rewards. We used a novel stimulus, the canoe movement detection task, which enabled us to observe the participants' decision thresholds directly. We used computational modeling to examine several aspects of the decision making process in these experiments. In what follows, we discuss our findings in each of these, separately.

\subsection{Allocation of limited time to decisions with different outcome}
In the previous studies that investigated the optimal speed-accuracy trade-off, all trials in each block were of the same type \cite{simen_reward_2009,balci_acquisition_2011,karsilar_speed_2014}. Our experimental design can be considered as an extension to these paradigms in that here the amount of time a participant should allocate to one type of decision to achieve the optimal performance, depends on the task parameters for all types of decisions. 

While we were writing this paper, another group of researchers, independently, reported their results on two experiments similar to what we reported here \cite{oud_irrational_2016}. Specifically, in study 2 in that paper, participants performed a perceptual decision making task in which the hard and easy trials were presented randomly in blocks with fixed duration. Also, the hard trials were associated with lower stakes than the easy trials. In both experiments, no cue was presented to the participants. Their results showed that the participants were slower in the hard trials than the easy trials. They concluded that the participants are spending too much time on hard trials which have low relative reward and so the behavior is sup-optimal. To provide stronger evidence for sub-optimality, they introduced an intervention in some blocks: in some randomly chosen trials of these blocks, the participants were motivated to make their choice faster. The results showed that in these intervention blocks, the participants achieved more rewards than the normal blocks, which shows that the participants are too slow when there is no intervention and so sub-optimal. 

Although these results show that the participants are sub-optimal, they do not necessarily show that the participants are spending too much time in only the hard trials. Since in both of their experiments the hard and easy trials are intermixed with no cue presented, the participants may have adopted a single time-decreasing optimal threshold. As we showed (see Figure \ref{fig:Simulate_B_WeibullThd}), in this case, although the behavior is optimal, the RT in the hard trials will be larger than the easy trials. Even if the participants adopt two different decision thresholds, and even if the threshold for the hard condition is lower than the easy condition, it is still possible to observe slower RT for hard trial just because the rate of information accumulation in the hard trials is lower.

Also, as the results of our study A showed, some participants are faster than optimal and some others are slower than optimal. The intervention introduced by \cite{oud_irrational_2016} can be useful for slower than optimal participants but will hurt the performance of the faster than optimal participants. They reported that in their perceptual decision making task, the intervention was beneficial for $60\%$ of the participants. Since in their experiment it was not possible to observe the decision thresholds directly, it is not possible to explain why this is the case.  

An interesting pattern that we observed in Experiment A, is that most of the participants used lower than optimal threshold (assuming that the threshold is time-constant) at the beginning of the experiment (see Figure \ref{fig:Results_A_SeperatedByInitiThd}). This means that these participants paid less attention to time than their accuracy. This is in contrast to some previous findings which showed that the participants used higher than optimal decision thresholds \cite{simen_reward_2009,balci_acquisition_2011}. The experiments used in these papers and in the current paper differ in several ways and so it is hard to decide why our results are not consistent with the results of these papers. Currently, we are conducting experiments similar to Experiment A but with the random dot motion as the stimulus to investigate this more. 

\subsection{Shape of decision thresholds}
We showed that in Experiment A, the optimal threshold for both the easy and hard trials is time-constant for most of the trial duration and decreases rapidly afterwards (left panel of Figure \ref{fig:OptimalThresholds}). To examine the shape of the decision thresholds that the participants adopted in this experiment, we fitted two versions of each computational model: one with time-constant thresholds and one with time-varying thresholds in which the threshold was modeled as a Weibull function. The results of comparing the models provided strong evidence favoring time-varying thresholds: for all participants and all models, the time-varying version of the model fitted better than the time-constant version. Most of the participants used time-decreasing boundaries with a shape different from the optimal thresholds (Figure 7 in S1 Text).

Time-decreasing thresholds are becoming more popular among researchers mostly because of two sources of evidence. First, recent neurophysiological findings support the notion of an ``urgency signal"  to make a decision as time elapses in a trial \cite{churchland_decision-making_2008}. This signal can be considered as being equivalent to time-decreasing thresholds \cite{thura_decision_2012}. Second, time-decreasing thresholds arise as the optimal solution in several experimental designs (e.g., deffered decision making with limited resourced to purchase information \cite{busemeyer_psychological_1988}, perceptual decision making with deadline \cite{frazier_sequential_2008}, perceptual decision making with mixed levels of difficulty and without cue \cite{drugowitsch_cost_2012}).

Despite this popularity, the behavioral evidence supporting these models is less prevalent. To the best of our knowledge, the most comprehensive comparison between the time-constant and time-decreasing models is provided by two recent papers by Hawkins et al. \cite{hawkins_revisiting_2015} and Voskuilen et al. \cite{voskuilen_comparing_2016}. Model comparison results of \cite{hawkins_revisiting_2015} showed that the time-constant thresholds were preferred over the time-decreasing thresholds for most of the participants. Consistent with these results, \cite{voskuilen_comparing_2016} found that the time-constant thresholds provide better fit to the data. In addition, the fitted time-varying thresholds were very similar to the time-constant thresholds because the amount of decrease in the threshold was very small.

The difference between our results and the results found by these papers could have several reasons. First, the stimuli used were different in these studies. \cite{hawkins_revisiting_2015} used random dot motion task, brightness discrimination task, and dot separation task. \cite{voskuilen_comparing_2016} used numerosity discrimination task and the random dot motion task. In all these experiments, the decision thresholds are not observable directly and their properties should be inferred from the choice and reaction time data. In contrast, in our canoe movement detection task, we inferred the shape of the decision threshold directly from the time series of the observed values of the decision thresholds in each trial. Another important difference is that the reaction time in the studies used in \cite{hawkins_revisiting_2015} and \cite{voskuilen_comparing_2016} is lower than in our experiment. For example, the median reaction time in all conditions of all 6 experiments reported in \cite{voskuilen_comparing_2016} is less than 0.8 secs, while the median reaction time in the easy condition of Experiments A of the current paper is more than 3 secs (middle panel of Figure \ref{fig:Results_A}). This difference is important because if the participants have to make their decisions very quickly, they may not have enough time to decrease their decision threshold. 

Second, in experiments reported in \cite{hawkins_revisiting_2015} and \cite{voskuilen_comparing_2016} the participants do not receive reward based on their performance. In contrast, in our experiments the participants are motivated to spend not too much time on each trial. Therefore, even in the easy trials, the participants needed less information to make their decisions as the time elapsed in a trial. Although, as we showed, this strategy was not optimal.    

\subsection{Optimal decision threshold}
We chose the parameters of Experiment A such that the optimal strategy for the hard trials was to respond as quickly as possible (zero threshold). Our statistical analysis showed that 12 out of 26 participants did not learn this simple strategy by the end of the experiment. Also, for the easy trials, some participants used higher than optimal and some others used lower than optimal decision thresholds. In Experiment B, the participants behaved sub-optimally by detecting the difficulty of the trial first and then setting the decision threshold. Sub-optimal behavior has been reported in some previous studies of the speed-accuracy trade-off. \cite{simen_reward_2009} found that the participants used higher than optimal decision thresholds. Also, the results of \cite{karsilar_speed_2014} showed that in experiments with deadline, where the optimal decision threshold is time-decreasing, the participants did not decrease their decision threshold within a trial. One reason for this sub-optimality could be the lack of enough practice. \cite{balci_acquisition_2011} showed that with extensive practice (about 14 sessions) the participants' decision threshold became closer to the optimal values. 

Every conclusion about the optimality should be made with caution. In all these studies, the optimality  is defined with respect to the actual value of the reward that a participant can earn. However, it is well-known that the subjective values of positive and negative rewards could be dramatically different from the actual values. If, for example, the absolute subjective value of negative rewards is larger than that of equal positive rewards, then the ``subjective optimal threshold" will be higher than the value of the decision threshold that maximizes the actual reward. In addition, we showed that in our experiments, the cost of time is equal to the average reward rate. The participants, however, may consider a subjective value for the cost of time. This will also change the shape and value of the optimal decision threshold. In a recent paper, Fudenberg et al. \cite{fudenberg_stochastic_2015} proved (theorem 6 in that paper) that for a diffusion process with arbitrary time-varying decision thresholds $\pm b(t)$, it is always possible to find a function $c(t)$ as the cost of time, such that $\pm b(t)$ is optimal in the sense that it leads to minimum value of the total cost. In other words, any observed decision threshold could be considered as optimal for a specific cost function.

\subsection{Adjusting decision threshold}
The main goal of this paper was to investigate how the participants adjust their decision threshold on a trial by trial basis. To this end, we proposed and compare several computational models for learning the decision thresholds. The results of the model comparison for Experiment A showed that the proposed RL model is the most likely model for most of the participants. However, based on the BIC of the fitted models, for some of the participants the model with no leaning (Model 1) and the model in which the learning is based only on the rewards (Model 3), provided better fit. Of course we do not expect that all participants use the same learning mechanism. The interesting result here is that although the RL model has a large number of parameters, it was the best model for many participants even based on BIC (which penalizes the complexity of a model more than AIC for large sample sizes).

Models 2-4 are based on the assumption that the participants adjust their decision threshold only based on the rewards they receive in each trial. Based on the values of the subjective rewards, these models can predict different patterns in the data. For example, if the absolute subjective value of the negative reward is smaller than the subjective value of the positive reward, these models predict that the participant reduces her decision threshold to zero for the hard trials in Experiment A. This is because the participant decreases her decision threshold after each correct trial and increases it after each incorrect trial. However, since the subjective value of positive rewards is higher, the amount of decrease is larger than increase and thus overall the decision threshold decreases. The results of the model comparison based on BIC, showed that one version of these models, Model 3, can explain the data of some of the participants better than other models.  

The heuristic models which assume the participants adjust their decision thresholds to achieve a desired level of accuracy (Models 5 and 6), or reaction time and accuracy (Model 7) did not win for any of the participants. Model 7 is an important competitor to the RL models. As we mentioned earlier, for an one-state SMDP this model is equivalent to the RL model.  Therefore, to be able to distinguish these models, we need at least a two-state experiment. Model 7 was inspired by the computational models of categorization \cite{nosofsky_attention_1986}, and multi-attribute decision making \cite{roe_multialternative_2001}. In this model, if we set $w_t=w_a=1$, the learning rule in Equation \ref{eq:A_model_acc_t_5} will reduce to $\Delta m_{s,k}=\alpha_a\cdot r_{k}+\alpha_t\cdot d_{k}+c_{s}$, where $r_k$ and $d_k$ are the reward and the time in trial $k$. This term resembles the TD error (Equation \ref{eq:RL_models_09}) in the RL models. The difference is that the last term for computing $\Delta m_{s,k}$ is constant, $c_s$, while in computing TD error we should use $\hat{V}_k(s')-\hat{V}_k(s)$. This term, which is the difference between the value of the current and the next state, is unique to the RL model. Model 7 is an instance of a ``supervised learning" algorithm: the desired values of the accuracy and reaction time are given and the participant should adjust her threshold to achieve these values. In contrast, in the RL model the desired levels are not known to the participant. In a sense, the state values determine the desired values. Therefore, in the RL models the participant both estimates the desired values and tries to reach those values simultaneously. 

\subsection{Effect of cue presentation}
Experiment B was designed to test the hypothesis that the participants set their decision threshold before the trial starts. In contrast to this hypothesis, the results of model comparison provided evidence for a model in which the participants first detect the difficulty of a trial and then set the corresponding decision threshold.

To the best of our knowledge, in all previous applications of the sequential sampling models, if trials with different levels of difficulty were intermixed, it was assumed that the participants use the same decision threshold for all trials, and the only parameter that varies between trials is the rate of information accumulation (drift rate) \cite{ratcliff_connectionist_1999,ratcliff_diffusion_2002,ratcliff_comparison_2004}. Again, we can think of two reasons for the inconsistency between the results of the previous studies and what we found in the current paper: first, short mean reaction times in these studies which does not give the participants enough time to first detect the difficulty of the stimulus, and second, since in the previous studies the participants are not rewarded based on their performance, they are not motivated to use different decision thresholds for different levels of difficulty.

\subsection{Limitations}
The experiments considered in this paper correspond to simple SMDPs. In particular, the transition probabilities are not a function of the actions taken by the participant. More research is necessary to investigate if the RL models can account for the participants data in more complicated environments \cite{hotaling_dynamic_2015}.

Another limitation of the current paper is that we have only applied our models to the data from the canoe task. It is interesting to investigate if the same results will be obtained if we use more conventional stimuli like the random dot motion task. The main issue is that since in these tasks the decision threshold is not observable directly, the learning models should be augmented with sequential sampling models. This makes the model fitting more complicated. We are currently working on this problem.

\textbf{Authors’ contributions.} A.K. and P.F. designed studies 1 and 2, conducted the statistical analyses and wrote the paper. J.B. conceived of the study, coordinated the study and helped draft the manuscript.  

\textbf{Funding.} The research was funded by the Air Force grant 067180-00002B. The opinions
expressed in this publication are those of the authors and do not
necessarily reflect the views of the funding agency.

\bibliography{references}

\end{document}